\documentclass[11pt]{article}
\usepackage{amsfonts,amssymb,amsmath}

\usepackage[english]{babel}
\newcommand{\halmos}{\rule{1ex}{1.4ex}}

\makeatletter
\@addtoreset{equation}{section}
\makeatother

\renewcommand{\theequation}{\thesection.\arabic{equation}}

\newtheorem{ittheorem}{Theorem}
\newtheorem{itlemma}{Lemma}
\newtheorem{itproposition}{Proposition}
\newtheorem{itdefinition}{Definition}
\newtheorem{itremark}{Remark}

\newenvironment{theorem}{\addtocounter{equation}{1}
\begin{ittheorem}}{\end{ittheorem}}

\newenvironment{lemma}{\addtocounter{equation}{1}
\begin{itlemma}}{\end{itlemma}}

\newenvironment{proposition}{\addtocounter{equation}{1}
\begin{itproposition}}{\end{itproposition}}

\newenvironment{definition}{\addtocounter{equation}{1}
\begin{itdefinition}}{\end{itdefinition}}

\newenvironment{remark}{\addtocounter{equation}{1}
\begin{itremark}}{\end{itremark}}

\newenvironment{proof}{\noindent {\em Proof}.\,\,\,}
{\hspace*{\fill}$\halmos$\medskip}

\newcommand{\beq}[1]{\begin{eqnarray}\label{#1}}
\newcommand{\eeq}{\end{eqnarray}}

\newcommand{\be}[1]{\begin{equation}\label{#1}}
\newcommand{\ee}{\end{equation}}

\newcommand{\bl}[1]{\begin{lemma}\label{#1}}
\newcommand{\el}{\end{lemma}}

\newcommand{\br}[1]{\begin{remark}\label{#1}}
\newcommand{\er}{\end{remark}}

\newcommand{\bt}[1]{\begin{theorem}\label{#1}}
\newcommand{\et}{\end{theorem}}

\newcommand{\bd}[1]{\begin{definition}\label{#1}}
\newcommand{\ed}{\end{definition}}

\newcommand{\bp}[1]{\begin{proposition}\label{#1}}
\newcommand{\ep}{\end{proposition}}

\newcommand{\bc}[1]{\begin{corollary}\label{#1}}
\newcommand{\ec}{\end{corollary}}

\newcommand{\bpr}{\begin{proof}}
\newcommand{\epr}{\end{proof}}

\newcommand{\bi}{\begin{itemize}}
\newcommand{\ei}{\end{itemize}}

\newcommand{\ben}{\begin{enumerate}}
\newcommand{\een}{\end{enumerate}}

\newcommand{\Z}{\mathbb Z}
\newcommand{\R}{\mathbb R}
\newcommand{\N}{\mathbb N}

\newcommand{\E}{\mathbb E}

\newcommand{\pee}{\mathbb P}
\newcommand{\bee}{\ensuremath{\mathcal{B}}}
\newcommand{\gee}{\ensuremath{\mathcal{G}}}

\newcommand{\ce}{\ensuremath{\mathcal{C}}}
\newcommand{\uu}{\ensuremath{\mathcal{U}}}
\newcommand{\s}{\ensuremath{\mathcal{S}}}
\newcommand{\loc}{\ensuremath{\mathcal{L}}}
\newcommand{\fe}{\ensuremath{\mathcal{F}}}

\newcommand{\vi}{\ensuremath{\varphi}}
\newcommand{\la}{\ensuremath{\Lambda}}
\newcommand{\si}{\ensuremath{\sigma}}

\begin{document}

\title{{\bf Possible loss
and recovery
of Gibbsianness
during\\
the stochastic evolution of Gibbs measures}}

\author{
A.C.D.\ van Enter
\footnote{Instituut voor Theoretische Natuurkunde, Rijksuniversiteit Groningen,
Nijenborg 4, 9747 AG Groningen, The Netherlands}\\
R.\ Fern\'andez
\footnote{Labo de Maths Raphael SALEM, UMR 6085, CNRS-Universit\'{e} de Rouen, Mathematiques, Site Colbert, F76821 Mont Saint Aignan, France}\\
F.\ den Hollander
\footnote{EURANDOM, Postbus 513, 5600 MB Eindhoven, The Netherlands}\\
F.\ Redig
\footnote{Faculteit Wiskunde en Informatica, Technische Universiteit Eindhoven, Postbus 513,
5600 MB Eindhoven, The Netherlands}
}

\maketitle

\footnotesize
\begin{quote}
{\bf Abstract:}
We consider Ising-spin systems starting from an initial Gibbs measure $\nu$
and evolving under a spin-flip dynamics towards a reversible Gibbs
measure $\mu\not=\nu$. Both $\nu$ and $\mu$ are assumed to have a finite-range
interaction. We study the Gibbsian character of the measure $\nu S(t)$ at time
$t$ and show the following:\\
(1) For all $\nu$ and $\mu$, $\nu S(t)$ is Gibbs for small $t$.\\
(2) If both $\nu$ and $\mu$ have a high or infinite temperature, then $\nu S(t)$
is Gibbs for all $t>0$.\\
(3) If $\nu$ has a low non-zero temperature and a zero magnetic field and $\mu$
has a high or infinite temperature, then $\nu S(t)$ is Gibbs for small $t$ and
non-Gibbs for large $t$.\\
(4) If $\nu$ has a low non-zero temperature and a non-zero magnetic field and
$\mu$ has a high or infinite temperature, then $\nu S(t)$ is Gibbs for small $t$,
non-Gibbs for intermediate $t$, and Gibbs for large $t$.\\
The regime where $\mu$ has a low or zero temperature and $t$ is not small
remains open. This regime presumably allows for many different scenarios.
\end{quote}
\normalsize

\vspace{12pt}


\section{Introduction}
\label{S1}

Changing interaction parameters, like the temperature or the magnetic field, in
a thermodynamical system is the preeminent way of studying such a system. In the
theory of interacting particle systems, which are used as microscopic models for thermodynamic systems, one associates
with each such interaction
parameter a class of stochastic evolutions, like Glauber dynamics or Kawasaki
dynamics.

In recent years there has been extensive interest in the {\it quenching regime},
in which one starts from a high- or infinite-temperature Gibbs state and considers
the behavior of the system under a low- or zero-temperature dynamics. This is
interpreted as a fast cooling procedure (which is different from the slow cooling
procedure of simulated annealing). One is interested in the asymptotic behavior
of the system, in particular, the occurrence of trapping in metastable frozen or
semi-frozen states (see \cite{FNI}, \cite{NS2}, \cite{NS3}, \cite{GNS}, \cite{NNS},
\cite{NS1},
\cite{CNS}).

Another regime that has been intensively studied is the one where, starting from a
low-non-zero-temperature Gibbs state of Ising spins in a positive magnetic field,
one considers a low-non-zero-temperature negative-magnetic-field Glauber dynamics
(see \cite{SchShl} and references therein). Under an appropriate rescaling of the
time and the magnetic-field strength, one finds a metastable transition from the
initial plus-state to the final minus-state.

In this paper we concentrate on the opposite case of the {\it unquenching regime},
in which one starts from a low-non-zero-temperature Gibbs state of Ising spins and
considers the behavior of the system under a high- or infinite-temperature Glauber
dynamics. This is interpreted as a fast heating procedure. As far as we know, this
regime has not been studied much (see e.g.\ \cite{asp}), as no singular behavior
was expected to occur. Although we indeed know that there is exponentially fast
convergence (cf.\ \cite{lig85}, Chapter 1, Theorem 4.1, and
\cite{MO1}, \cite{MO2}) to the high- or infinite-temperature Gibbs
state (i.e., the asymptotic behavior is unproblematic), we will show that {\it
at sharp finite times there can be transitions between regimes where the evolved
state is Gibbsian and regimes where the evolved state is non-Gibbsian}.

In the light of the results in \cite{Enter}, Chapter 4, on renormalization-group transformations,
it should perhaps not come as a surprise that such transitions can happen. Indeed, we
can view the time-evolved measure as a kind of
(single-site)
renormalized Gibbs measure. Even though
the image spin $\si_t (x)$ at time $t$ at site $x$ is not a
(random)
function of the original spins $\sigma_0 (y)$ at time $0$
for $y$ in only a finite block around $x$, by the Feller character of the Glauber dynamics
it depends only weakly on the spins $\sigma_0 (y)$ with $y$ large. In that sense the
time evolution is close to a standard renormalization-group transformation,
without
rescaling, and so we can expect Griffiths-Pearce
pathologies.

We will prove the following:
\begin{enumerate}
\item[(1)]
For an arbitrary initial Gibbs measure and an arbitrary Glauber dynamics, both having
finite range, the measure stays Gibbs in a small time interval, whose length depends on
both the initial measure and the dynamics (Theorem \ref{Gibbscons}). This result, though
somewhat surprising, essentially comes from the fact that for small times the set of
sites where a spin flip has occurred consists of ``small islands'' that are far apart
in a ``sea'' of sites where no spin flip has occurred.
\item[(2)]
For a high- or infinite-temperature initial Gibbs measure and a high- or
infinite-tempera\-ture Glauber dynamics, the measure is Gibbs for all $t>0$
(Theorems \ref{Dobr} and \ref{Dobr*}).
\item[(3)]
For a low-non-zero-temperature initial Gibbs measure and a high- or infinite-temperature
Glauber dynamics, there is a transition from Gibbs to non-Gibbs (Theorems
5.16 and 6.18). This result is somewhat
counter-intuitive: after some time of heating the system it reaches a high temperature,
where a priori we would expect the measure to be well-behaved because it should be
exponentially close to a
Completely Analytic (see \cite{DS})
high-temperature Gibbs measure. As we will see, this intuition is wrong.
However, from the results of \cite{ MV94} it follows that this transition does not occur
when the initial measure is a rigid ground state (zero-temperature) measure
(i.e., a Dirac measure).

\item[(4)]
For a low-non-zero-temperature initial Gibbs measure and a high- or infinite-temperature
Glau\-ber dynamics, there possibly is a transition back from non-Gibbs to Gibbs when
the Hamiltonian of
the initial Gibbs measure has a non-zero magnetic field (Theorems \ref{Isingthm} and \ref{Isingthm*}).
\end{enumerate}
The complementary regimes, with a low- or zero-temperature Glauber dynamics acting over
large times, are left open.

In Section \ref{S2} we start by giving some basic notations and definitions, and
formulating some general facts.

In Section \ref{S3} we give representations of the conditional probabilities of the
time-evolved measure and clarify the link between the Gibbsian character of the
time-evolved measure and the Feller property of the backwards process. These results
are useful for proving the ``positive side'', i.e., for showing that the time-evolved
measure is Gibbsian. We use a criterion of \cite{Enter}, Chapter 4, Step 1, or \cite{FerPfis} to
identify {\it bad configurations} (points of essential discontinuity of every
version of the conditional probabilities) as those configurations for which the
{\it constrained} system (i.e., the measure at time $0$ conditioned on the
{\em future}
bad
configuration at time $t>0$) exhibits a phase transition. This criterion will
serve for the ``negative side'', i.e., for showing that the time-evolved measure is
non-Gibbsian.

In Section \ref{S4} we prove that for an arbitrary initial measure and an arbitrary
dynamics, both having finite-range interactions, the measure at time $t$ is Gibbs
for all $t \in [0,t_0]$, where $t_0$ depends on the interactions.

In Section \ref{S5} we treat the case of infinite-temperature dynamics, i.e., a
product of independent Markov chains. This example already exhibits all the
transitions between Gibbs and non-Gibbs we are after. Moreover, it has the advantage of
fitting exactly in the framework of the renormalization-group transformations:
the time-evolved measure is nothing but a single-site Kadanoff transform
of the original measure, where the parameter $p(t)$ of this transform
varies continuously from $p(0)=\infty$ to $p(\infty)=0$. For the case of a
low-temperature initial measure we restrict ourselves to the $d$-dimensional
Ising model.

In Section \ref{S6} we show that the results of Section \ref{S5} also apply in the
case of a high-temperature dynamics. The basic ingredient is a cluster expansion in
space and time, as developed in \cite{MV93} and worked out in detail in \cite{MN}.
This is
formulated in Theorem \ref{clust} and is the technical tool needed to develop the
``perturbation theory'' around the infinite-temperature case.

In Section \ref{S7} we give a dynamical interpretation of the transition from Gibbs
to non-Gibbs in terms of a change in the {\it most probable history of an
improbable configuration}. We show that the transition is not linked with a wrong behavior
in the large deviations at fixed time, and we close by formulating a number
of open problems.


\section{ Notations and definitions}
\label{S2}

\subsection{Configuration space}
\label{S2.1}

The configuration space of our system is $\Omega = \{-1,+1\}^{\Z^d}$, endowed with the
product topology. Elements of $\Omega$ are denoted by $\sigma,\eta$. A configuration
$\sigma$ assigns to each lattice point $x\in \Z^d$ a spin value $\sigma (x) \in
\{-1,+1\}$. The set of all finite subsets of $\Z^d$ is denoted by $\mathcal {S}$.
For $\Lambda\in \mathcal{S}$ and $\sigma\in \Omega$, we denote by $\sigma_\Lambda$ the
restriction of $\sigma$ to $\Lambda$, while $\Omega_\Lambda$ denotes the set of all
such restrictions. A function $f:\Omega\rightarrow\R$ is called local if there
exists a finite set $\Delta \subset \Z^d$ such that $f(\eta) = f(\sigma)$ for $\sigma$
and $\eta$ coinciding on $\Delta$. The minimal such $\Delta$ is called the dependence
set of $f$ and is denoted by $D_f$. The vector space of all local functions is denoted
by $\loc$. This is a uniformly dense subalgebra of the set of all continuous functions
$\mathcal{C}(\Omega )$. A local function $f:\Omega\rightarrow\R$ with dependence set
$D_f\subset\la$ can be viewed as a function on $\Omega_\la$. With a slight abuse of
notation we use $f$ for both objects. For $\sigma,\eta\in \Omega$ and $\Lambda\subset
\Z^d$, we denote by $\sigma_\Lambda\eta_{\Lambda^c}$ the configuration whose restriction
to $\Lambda$ (resp.\ $\Lambda^c$) coincides with $\sigma_\Lambda$ (resp.\ $\eta_{\Lambda^c}$).
For $x\in \Z^d$ and $\sigma\in\Omega$, we denote by $\tau_x\sigma$ the shifted configuration
defined by $\tau_x\sigma (y) = \sigma (y+x)$. A sequence of probability measures $\mu_\la$
on $\Omega_\la$ is said to converge to a probability measure $\mu$ on $\Omega$ (notation
$\mu_\la\rightarrow\mu$) if
\be{limiet}
\lim_{\la\uparrow\Z^d} \int f \ d\mu_\la = \int f \ d\mu \qquad \forall f\in\loc.
\ee

\subsection{Dynamics}
\label{S2.2}

The dynamics we consider in this paper is governed by a collection of spin-flip rates
$c(x,\sigma)$, $x\in\Z^d,\sigma\in\Omega$, satisfying the following conditions:
\begin{enumerate}
\label{rates}
\item
{\it Finite range}: $c_x:\sigma\mapsto c(x,\sigma)$ is a local function of
$\sigma$ for all $x$, with $\mbox{diam}(D_{c_x})\leq R<\infty$.
\item
{\it Translation invariance}: $\tau_x c_0= c_x$ for all $x$.
\item
{\it
Strict positivity
}: $c(x,\sigma) >0$ for all $x$ and $\sigma$.
\end{enumerate}
Note that these conditions imply that there exist $\epsilon, M\in
(0,\infty)$ such that
\be{posrate}
0 < \epsilon\leq c(x,\sigma) \leq M <\infty \qquad \forall x\in\Z^d,\sigma\in\Omega.
\ee
Given the rates $(c_x)$, we consider the generator defined by
\be{geny}
Lf = \sum_{x\in\Z^d} c_x \nabla_x f \qquad \forall f\in\loc,
\ee
where
\be{nabla}
\nabla_x f (\sigma ) = f(\sigma^x) -f(\sigma).
\ee
Here, $\sigma^x$ denotes the configuration defined by $\sigma^x (x)= -\sigma (x)$ and
$\sigma^x (y)= \sigma (y)$ for $y\not= x$. In \cite{lig85}, Theorem 3.9,
it is proved that the closure
of $L$
on
$\mathcal{C}(\Omega)$ is the generator of a unique Feller process $\{\sigma_t:
t\geq 0 \}$. We denote by $S(t)=\exp (tL)$ the corresponding semigroup, by $\pee_\sigma$
the path-space measure given $\sigma_0=\sigma$, and by $\E_\sigma$ expectation over
$\pee_\sigma$.

A probability measure $\mu$ on the Borel $\si$-field of $\Omega$ is called {\it invariant}
if
\be{intL}
\int Lf d\mu =0 \qquad \forall f\in\loc.
\ee
It is called {\it reversible} if
\be{Lfg}
\int (Lf)g d\mu = \int f (Lg) d\mu \qquad \forall f,g\in\loc.
\ee
Reversibility implies invariance. For spin-flip dynamics with generator $L$ defined by
(\ref{geny}), reversibility of $\mu$ is equivalent to
\be{detbal}
c(x,\sigma^x) \frac{d\mu^x}{d\mu} = c(x,\sigma) \qquad \forall x\in\Z^d,\sigma\in\Omega,
\ee
where $\mu^x$ denotes the distribution of $\sigma^x$ when $\sigma$ is distributed
according to $\mu$. Note that (\ref{detbal}) implies the existence of a continuous
version of the
Radon-Nikod\'ym (RN)-derivative $d\mu^x/d\mu$. This will be important in the sequel.

\subsection{Interactions and Gibbs measures}
\label{S2.3}

A {\it good} interaction is a function
\be{U}
U:\s\times\Omega\rightarrow\R,
\ee
such that the following two conditions are satisfied:
\begin{enumerate}
\item
{\it Local potentials in the interaction:} $U(A,\sigma)$ depends on $\sigma (x),
x\in A$, only.
\item
{\it Uniform summability:}
\be{sumA}
\sum_{A\ni x} \sup_{\sigma\in \Omega} | U(A,\sigma )| <\infty \qquad x\in\Z^d.
\ee
\end{enumerate}
The set of all good interactions will be denoted by $\bee$. A good interaction is called
{\it translation invariant} if
\be{UAx}
U(A+x,\tau_{-x}\sigma)= U(A,\sigma ) \qquad \forall A\in\s, x\in\Z^d, \sigma\in \Omega.
\ee
The set of all translation-invariant good interactions is denoted by $\bee_{ti}$.
An interaction $U$ is called {\it finite-range} if there exists an $R>0$ such that
$U(A,\sigma )=0$ for all $A\in \s$ with $\mbox{diam}(A)>R$. The set of all finite-range
interactions is denoted by $\bee^{fr}$ and the set of all translation-invariant finite-range
interactions by $\bee^{fr}_{ti}$. For $U\in \bee$, $\zeta\in\Omega$, $\Lambda\in\s$, we define the finite-volume Hamiltonian with boundary condition $\zeta$ as
\be{Hzeta}
H^\zeta_\Lambda (\sigma )= \sum_{A\cap\Lambda\not=\emptyset}
U(A,\sigma_\Lambda\zeta_{\Lambda^c})
\ee
and the Hamiltonian with free boundary condition as
\be{HLambda}
H_\Lambda (\sigma ) = \sum_{A\subset \Lambda} U(A,\sigma),
\ee
which depends only on the spins inside $\Lambda$. Corresponding to the Hamiltonian in
(\ref{Hzeta}) we have the finite-volume Gibbs measures $\mu^{U,\zeta}_\Lambda$,
$\Lambda\in\s$, defined on $\Omega$ by
\be{finvol}
\int f(\xi) \mu_\Lambda^{U,\zeta} (d\xi)
= \sum_{\sigma_\Lambda\in\Omega_\Lambda} f(\sigma_\Lambda\zeta_{\Lambda^c})
\frac{\exp{[-H^\zeta_\Lambda (\sigma)]}}{Z^\zeta_\Lambda},
\ee
where $Z_\Lambda^\zeta$ denotes the partition function normalizing $\mu^{U,\zeta}_\Lambda$ to
a probability measure.

For a probability measure $\mu$ on $\Omega$, we denote by $\mu^\zeta_\Lambda$ the
conditional probability distribution of $\sigma (x), x\in \Lambda$, given
$\sigma_{\Lambda^c}=\zeta_{\Lambda^c}$. Of course, this object is only defined on
a set of $\mu$-measure one. For $\Lambda\in\s, \Gamma\in \s$ and $\Lambda\subset\Gamma$,
we denote by $\mu_\Gamma(\sigma_\Lambda|\zeta)$ the conditional probability to find
$\sigma_\Lambda$ inside $\Lambda$, given that $\zeta$ occurs on $\Gamma\setminus\Lambda$.
For $U\in \bee$, we call $\mu$ a Gibbs measure with interaction $U$ if its conditional
probabilities coincide with the ones prescribed in (\ref{finvol}), i.e., if
\be{gibbsdef}
\mu^\zeta_\Lambda = \mu^{U,\zeta}_\Lambda \qquad \mu-a.s. \qquad \Lambda\in\s,
\zeta\in\Omega.
\ee
We denote by $\mathcal{G} (U)$ the set of all Gibbs measures with interaction $U$. For
any $U\in \bee$, $\gee (U)$ is a non-empty compact convex set. The set of all Gibbs
measures is
\be{gee}
\gee= \bigcup_{U\in \bee} \gee (U).
\ee
Note that $\gee$ is not a convex set, since for $U$ and $V$ in
$\bee_{ti}$, convex combinations of $\gee (U)$ and $\gee (V)$
are not in $\gee$ unless $\gee(U)=\gee (V)$ (see \cite{Enter} section 4.5.1).\\
{\bf Remark:} We will often use the notation $H=\sum_{A} U(A,\cdot)$ for
the ``Hamiltonian" corresponding to the interaction $U$. This formal
sum has to be interpreted as the collection of ``energy differences",
i.e., if $\si$ and $\eta$ agree outside a finite volume $\la$, then:
\begin{equation}\label{formalsum}
H(\eta) -H(\si ) = \sum_{A\cap \Lambda \not=\emptyset} [U(A,\eta) -U(A,\si)].
\end{equation}

\bd{Gibbsdef}
A measure $\mu$ is called {\bf Gibbsian} if $\mu\in\gee$, otherwise it is called
{\bf non-Gibbsian}.
\ed

\subsection{Gibbsian and non-Gibbsian measures}
\label{S2.4}

In this paper we study the time-dependence of the Gibbsian property of a measure under
the stochastic evolution $S(t)$. In other words, we want to investigate whether or not
$\nu S(t)\in\gee$ at a given time $t>0$.

\bp{Kozlov}
The following three statements are equivalent:
\begin{enumerate}
\item
$\mu\in\gee$.
\item
$\mu$ admits a continuous and strictly positive version of its conditional probabilities
$\mu^\zeta_\Lambda$, $\Lambda\in\s,\zeta\in\Omega$.
\item
$\mu$ admits a continuous version of the RN-derivatives $d\mu^x/d\mu$, $x\in \Z^d$.
\end{enumerate}
\ep

\bpr
See \cite{Koz} and \cite{Sul}.
\epr

\noindent
We will mainly use item 3 and look for a continuous version of the RN-derivatives
$d\mu^x/d\mu$ by approximating them uniformly with local functions.

A necessary and sufficient condition for $\mu$ not to be Gibbsian ($\mu\not\in\gee$)
is the existence of a bad configuration, i.e., a point of essential discontinuity.
This is defined as follows:
\bd{baddef}
A configuration $\eta\in \Omega$ is called {\bf bad} for a probability measure $\mu$
if there exists $\epsilon >0$ and $x\in\Z^d$ such that for all $\Lambda\in \s$ there exist $\Gamma \supset\Lambda$ and $\xi,\zeta\in\Omega$ such that:
\be{muGamma}
\left|\mu_\Gamma (\sigma(x)|\eta_{\Lambda\setminus\{x\}}\zeta_{\Gamma\setminus\Lambda})
- \mu_\Gamma (\sigma(x)|\eta_{\Lambda\setminus\{x\}}\xi_{\Gamma\setminus\Lambda})
\right|>\epsilon.
\ee
\ed
Note that in this definition only the finite-dimensional distributions of
$\mu$ enter. It is clear that a bad configuration is a point of discontinuity
of {\it every} version of the conditional probabilities of $\mu$. Conversely,
a measure that has no bad configurations is Gibbsian (see e.g.\ \cite{MRV}).

\subsection{Main question}
\label{S2.5}

Our starting points in this paper are the following ingredients:
\begin{enumerate}
\item
{\bf
A translation invariant
initial measure $\nu\in \gee (U_\nu)$}, corresponding to a finite-range
translation-invariant interaction $U_\nu \in \bee^{fr}_{ti}$ as introduced in
Section \ref{S2.3}.
\item
{\bf A spin-flip dynamics}, with flip rates as introduced in Section \ref{S2.2}.
This dynamics has a {\it reversible} measure $\mu$, which
satisfies
\be{dmuxdmu1}
\frac{d\mu^x}{d\mu} = \frac{c(x,\sigma)}{c(x,\sigma^x)}.
\ee
Hence, by Proposition \ref{Kozlov} there exists an interaction $U_\mu\in\bee$
such that $\mu\in \gee (U_\mu)$. Since the rates are translation invariant
and have finite range, this interaction can actually be chosen in $\bee^{fr}_{ti}$
and satisfies (recall (\ref{Hzeta}) and (\ref{gibbsdef}))
\be{dmuxdmu2}
\frac{d\mu^x}{d\mu}= \exp \left(\sum_{A\ni x} [U_\mu(A,\sigma)-U_\mu (A,\sigma^x)]\right).
\ee
Without loss of generality we can take the rates $c(x,\sigma)$ of the form
\be{cx}
c(x,\sigma) =\exp\left(\frac{1}{2}\sum_{A\ni x} [U_\mu(A,\sigma)-U_\mu (A,\sigma^x)]\right).
\ee
\end{enumerate}

\noindent
A finite-volume approximation of the rates in (\ref{cx}) that we will often use is given by
\be{aprate}
c_\Lambda (x,\sigma) =\exp \left[ H^\mu_\Lambda (\sigma)- H^\mu_\Lambda (\sigma^x)\right],
\ee
where $H^\mu_\Lambda$ is the Hamiltonian with free boundary condition associated with the
interaction $U_\mu$ (recall (\ref{HLambda})). These rates generate a pure-jump process
on $\Omega_\Lambda =\{-1,+1\}^\Lambda$ with generator
\be{finvolgen}
(L_\Lambda f)(\cdot) = \sum_{x\in \Lambda} c_\Lambda (x,\cdot)\nabla_x f(\cdot)
\qquad \forall f\in\loc.
\ee
Since $L_\Lambda f$ converges to $Lf$ as $\Lambda\uparrow\Z^d$ for any local function
$f\in \loc$, the corresponding semigroup $S_\Lambda (t)$ converges strongly in the
uniform topology on $\ce (\Omega)$ to the semigroup $S(t)$, i.e., $S_\Lambda (t) f
\rightarrow S(t) f$ as $\Lambda\uparrow\Z^d$ in the uniform topology for any
$f\in C(\Omega)$. Therefore we have the following useful approximation result. Let
$\nu$ be a probability measure on $\Omega$ and $\nu_\Lambda$ its restriction to
$\Omega_\Lambda$ (viewed as a subset of $\Omega$). Then
\be{approx}
\lim_{\Lambda\uparrow\Z^d} \nu_\Lambda S_\Lambda (t) = \nu S(t),
\ee
where the limit is in the sense of (\ref{limiet}). If $\nu\in\gee (U_\nu)$ is a Gibbs
measure, then we can replace the finite-volume restriction $\nu_\Lambda$ by the
free-boundary-condition finite-volume Gibbs measure (in the case of no phase transition),
or by the appropriate finite-volume Gibbs measure with
generalized boundary condition that approximates $\nu$ (in the case
of a phase transition).

The main question that we will address in this paper is the following:\\
\begin{flushleft}
{\bf Question:}
\end{flushleft}
\begin{center}
Is $\nu S(t)=\nu_t$ a Gibbs measure?
\end{center}
In order to study this rather general question we have to distinguish between different
regimes, as defined next.

\bd{hightemp}
$U\in\bee$ is a {\it high-temperature} interaction if
\be{DobU}
\sup_{x\in\Z^d} \sum_{A\ni x}
(|A|-1)\sup_{\sigma, \sigma'\in\Omega}|U(A,\sigma)-U(A,\sigma')| < 2.
\ee
\ed
Equation (\ref{DobU}) implies the Dobrushin uniqueness condition for the associated
conditional probabilities $\mu^{U,\zeta}_\Lambda$, $\Lambda\in\s,\zeta\in\Omega$.
In particular, it implies that $|\gee (U)|=1$ (i.e., no phase transition). Note that
it is independent of the ``single-site part'' of the interaction, i.e., of the
interactions $U(\{x\},\sigma)$.\\
{\bf Remark:}
We interpret the above norm as an inverse temperature, so small norm
means high temperature.

\bd{dyninit}
We call:
\begin{enumerate}
\item
an initial measure $\nu$ ``high-temperature'' if it has an interaction
satisfying (\ref{DobU}), and write $T_\nu >>1$.
\item
an initial measure $\nu$ ``infinite-temperature'' if it is a product measure,
(i.e., if the corresponding interaction $U_\nu$ satisfies $U_\nu(A,\sigma)=0$
for all $A$ with $|A|>1$), and write $T_\nu =\infty$.
\item
a dynamics ``high-temperature'' if the associated reversible Gibbs measure
$\mu$ has an interaction $U_\mu$ satisfying (\ref{DobU}), and write $T_\mu >>1$.
\item
a dynamics ``infinite-temperature'' if the associated reversible measure $\mu$
is a product measure (i.e., if the corresponding interaction $U_\mu$ satisfies
$U_\mu(A,\sigma)=0$ for all $A$ with $|A|>1$), and write $T_\mu=\infty$.
\end{enumerate}
\ed

\noindent
As we will see in Section \ref{S5}, the study of infinite-temperature dynamics is
particularly instructive, since it can be treated essentially completely and already
contains all the interesting phenomena we are after.


\section{General facts}
\label{S3}

\subsection{Representation of the RN-derivative}
\label{S3.1}

As summarized in Proposition \ref{Kozlov}, an object of particular use in the investigation
of the Gibbsian character of a measure is its RN-derivative $d\mu^x/d\mu$ w.r.t.\ a spin
flip at site $x$. In this section we show how to exploit the reversibility of the dynamics
in order to obtain a sequence of continuous functions converging to the RN-derivative of
the time-evolved measure $\nu_t = \nu S(t)$ w.r.t.\ spin flip. Let us first consider
the finite-volume case. We start from the finite-volume generator
\be{LLambda}
L_\Lambda f(\sigma ) =\sum_{x\in \Lambda} c_\Lambda (x,\sigma ) (f(\sigma^x)
-f(\sigma)),
\ee
where the finite-volume rates $c_\Lambda (x,\cdot)$ are given by (\ref{aprate}). Suppose
that our starting measure $\nu\in \gee (U_\nu )$ is such that $|\gee (U_\nu )|=1$, which
implies that the free-boundary-condition finite-volume approximations $\nu_\Lambda$ converge
to $\nu$. The free-boundary-condition finite-volume Gibbs measure $\mu_\Lambda$,
corresponding to the interaction $U_\mu$, is the reversible measure of the generator
$L_\Lambda$. We can then compute, using reversibility,
\beq{revRN}
\frac {d\nu_\Lambda S_\Lambda(t)^x}{d\nu_\Lambda S_\Lambda(t)} (\sigma )
&=&
\left(\frac {d\nu_\la S_\la(t)^x}{d\mu_\la S_\la(t)^x} (\sigma )\right)
\left(\frac {d\mu_\la S_\la(t)^x}{d\mu_\la S_\la(t)} (\sigma )\right)
\left(\frac {d\mu_\la S_\la(t)}{d\nu_\la S_\la(t)} (\sigma )\right)
\nonumber\\
&=&
\left(\frac {d\nu_\la S_\la(t)}{d\mu_\la S_\la(t)} (\sigma^x )\right)
\left(\frac {d\mu_\la^x}{d\mu_\la} (\sigma )\right)
\left(\frac {d\mu_\la S_\la(t)}{d\nu_\la S_\la(t)} (\sigma )\right)
\nonumber\\
&=&
\left[S_\la (t)\left(\frac {d\nu_\la}{d\mu_\la }\right) (\sigma^x )\right]
\left(\frac {d\mu_\la^x}{d\mu_\la} (\sigma )\right)
\left[S_\la (t)\left(\frac {d\nu_\la}{d\mu_\la }\right) (\si )\right]^{-1}.
\eeq

\bd{difham}
$H^{\mu,\nu}_\Lambda (\sigma)= \sum_{A\subset\la} [U_\mu(A,\sigma)-U_\nu (A,\sigma)]$.
Note that this ``difference Hamiltonian'' depends on both the initial measure
and the dynamics.
\ed

\noindent
Using this definition, we may rewrite (\ref{revRN}) as
\be{haha}
\frac {d\nu_\Lambda S_\Lambda(t)^x}{d\nu_\Lambda S_\Lambda(t)} (\sigma )
= \frac{d\mu^x_\Lambda}{d\mu_\Lambda} (\sigma)
\frac{\E_{\sigma^x}^\la \left( \exp [H^{\mu,\nu}_\la (\sigma_t)]\right)}
{\E^\la_\sigma \left(\exp [H^{\mu,\nu}_\la (\sigma_t)]\right)},
\ee
where $\E^\la_\sigma$
denotes the
expectation for the process with semigroup $S_\la (t)$
starting from $\sigma$. Since this semigroup converges to the semigroup $S(t)$
of the infinite-volume process as $\la\to\Z^d$, we obtain the following:

\bp{prop 1}
For any $\si\in \Omega$ and $t\geq 0$,
\be{revprop}
\frac{d\nu S(t)^x}{d\nu S(t)} (\sigma)
= \frac{d\mu^x}{d\mu} (\sigma )
\lim_{\la\uparrow \Z^d}
\frac{\E_{\sigma^x} \left( \exp [H^{\mu,\nu}_\la (\sigma_t)]\right)}
{\E_\sigma \left(\exp [H^{\mu,\nu}_\la (\sigma_t)]\right)},
\ee
where this equality is to be interpreted as follows:
if the limit in the RHS of (\ref{revprop}) is a limit in the uniform
topology, then it defines a continuous version of the LHS.
\ep

\bpr
The claim follows from a combination of (\ref{approx}) and
(\ref{haha}) with Lemma \ref{lconv} below.
\epr

\bl{lconv}
If $\nu_n\rightarrow\nu$ weakly as $n\to\infty$, and $d\nu_n^x/d\nu_n\in \ce(\Omega)$
exists for any $n\in \N$ and converges uniformly to a continuous function $\Psi$, then
\be{Psi}
\Psi=\lim_{n\uparrow\infty} \frac{d\nu_n^x}{d\nu_n}= \frac{d\nu^x}{d\nu}.
\ee
\el

\bpr
Let $f:\Omega\rightarrow\R$ be a continuous function. Define
$\theta_x :\Omega\to\Omega$ by $\theta_x (\si) =\si^x$.
Then also
$f\circ\theta_x:\Omega\rightarrow\R$ is a continuous function. Therefore
\beq{triv}
\int f d\nu^x &=&
\int \left(f\circ \theta_x (\sigma )\right)\nu (d\sigma )\nonumber\\
&=&
\lim_{n\uparrow\infty}
\int \left(f\circ\theta_x (\sigma )\right)\nu_n (d\sigma )\nonumber\\
&=&
\lim_{n\uparrow\infty}
\int \frac{d\nu_n^x}{d\nu_n} (\sigma ) f(\sigma )\nu_n (d\sigma )
\nonumber\\
&=&
\lim_{n\uparrow\infty}\int \Psi(\sigma) f(\sigma) \nu_n(d\sigma)
\nonumber\\
&=& \int \Psi f d\nu,
\eeq
where the fourth equality follows from
\be{limfrac}
\lim_{n\uparrow\infty}
\int \Big|\frac{d\nu_n^x}{d\nu_n} (\sigma )-\Psi (\sigma )\Big|f(\sigma)\nu_n (d\sigma)
\leq  \lim_{n\uparrow\infty}
\| f\|_\infty \| \frac{d\nu_n^x}{d\nu_n}-\Psi \|_\infty =0.
\ee
Since (\ref{triv}) holds for any continuous function $f$, the statement of the
lemma follows from the Riesz representation theorem.
\epr

Proposition \ref{prop 1}, combined with Proposition \ref{Kozlov}, will be used
in Sections \ref{S4}--\ref{S6} to prove Gibbsianness.

\subsection{Path-space representation of the RN-derivative}
\label{S3.2}

An alternative representation of the RN-derivative $d\nu_t^x/d\nu_t$ is
obtained by observing that $\nu_t=\nu S(t)$ is the restriction of the path-space measure
$\pee_\nu^{[0,t]}$ to the ``layer" $\{t\}\times\Omega$. In some sense, this
path-space measure can be given a Gibbsian representation with the help of
Girsanov's formula. The ``relative energy for spin flip" of this path-space
measure is a well-defined (though unbounded) random variable. Conditioning the path-space
measure RN-derivative for a spin flip at site $x$ in the layer $\{t\}\times\Omega $,
we get the RN-derivative $d\nu_t^x/d\nu_t$. More formally, let us denote by $\pi_t$
the projection on time $t$ in path space, i.e., $\pi_t (\omega )= \omega_t$ with
$\omega\in D([0,t],\Omega)$ the Skorokhod space. By a spin flip at site $x$ in path
space we mean a transformation
\be{pathflip}
\Theta_x : D([0,t],\Omega)\rightarrow D([0,t],\Omega )
\ee
such that
\be{pathflip'}
(\pi_t(\omega))^x = \pi_t (\Theta_x(\omega)).
\ee
Different choices are possible, but in this section we choose
\be{Thetax}
(\Theta_x (\omega))(s,y) =
\left\{\begin{array}{ll}
-\omega (s,x) &\mbox{for} \ y=x, \ 0\leq s\leq t,\\
\omega (s,y) &\mbox{otherwise}.
\end{array}
\right.
\ee
Let $\fe_{[t]}$ denote the $\sigma$-field generated by the projection $\pi_t$. Then we
can write the following formula:
\be{pathRNd}
\frac{d\nu S(t)^x}{d\nu S(t)} = \E_\nu^{[0,t]} \left(
\frac{d\pee_\nu^{[0,t]}\circ\Theta_x}{ d\pee_\nu^{[0,t]}}\mid
\fe_{[t]}\right).
\ee
This equation is useful because of the Gibbsian form of the RHS of (\ref{pathRNd})
given by Girsanov's formula, as shown in the proof of the following:

\bp{prop 2}
Let $\nu$ be a Gibbs measure on $\Omega$. For any $t>0$,
\be{abscont}
\nu S(t)^x << \nu S(t)
\ee
and the RN-derivative can be written in the form
\be{pathspaceRN}
\frac{d\nu S(t)^x}{d\nu S(t)}=
\E_\nu^{[0,t]}\left[
\left(\frac{d\nu^x}{d\nu}\circ\pi_0\right)\Psi_x\mid\fe_{[t]}\right],
\ee
where $\Psi_x: D([0,t],\Omega)\rightarrow \R$ is a continuous function
on path space (in the Skorokhod topology).
\ep

\bpr
We first approximate our process by finite-volume pure-jump processes
and use Girsanov's formula to obtain the densities of these processes w.r.t.\ the
independent spin-flip process. Indeed, denote by $\pee_\sigma^\la$
the path-space measure of the finite-volume approximation with generator (\ref{finvolgen})
and by $\pee_\sigma^{\la ,0}$ the path-space measure of the independent spin-flip process
in $\la$, i.e., the process with generator
\be{L0f}
L^0_\la f = \sum_{x\in \la}\nabla_x f \qquad f\in\loc.
\ee
We have for $f: \Omega\rightarrow\R$ such that $D_f\subset\la$,
\beq{pathblub}
\int f(\sigma )\ \nu S(t)^x (d\sigma )
&=&
\lim_{\la\uparrow\Z^d}\int \nu (d\si )\int \pee^\la_\si (d\omega)
\,f\left(\pi_t\left(\Theta_x(\omega)\right)\right)\nonumber\\
&=&
\lim_{\la\uparrow\Z^d}\int \nu (d\si )
\int \pee^{\la,0}_\si (d\omega)
\frac{d\pee^\la_\si}{d\pee^{\la,0}_\si}(\omega)
\,f\left(\pi_t\left(\Theta_x(\omega)\right)\right).
\eeq
Since $\pee^{\la,0}_\si$ is the path-space measure of the independent
spin-flip process, the transformed measure $\pee^{\la,0}_\si\circ\Theta_x$ equals
$\pee^{\la,0}_{\si^x}$. Abbreviate
\be{Fla}
F_\la (\omega) = \frac{d\pee^\la_{\omega_0}}{d\pee^{\la,0}_{\omega_0}}(\omega).
\ee
Then we obtain
\beq{intnudsi}
&&\int \nu (d\si )\int \pee^{\la,0}_\si (d\omega) F_\la (\omega)
f(\pi_t (\Theta_x(\omega)))\nonumber\\
&&\qquad\qquad =
\int \nu (d\sigma )
\int \pee^{\la,0}_{\si^x} (d\omega) F_\la (\Theta_x (\omega))f(\pi_t(\omega))\nonumber\\
&&\qquad\qquad =
\int \nu (d\sigma )\int \pee_{\si^x}^\la (d\omega)
\frac{d\pee^{\la,0}_{\si^x}}{d\pee^\la_{\si^x}}(\omega)
F_\la (\Theta_x (\omega)) f(\pi_t(\omega))\nonumber\\
&&\qquad\qquad =
\int \nu (d\sigma) \frac{d\nu^x}{d\nu} (\sigma )\int \pee^\la_\si (d\omega)
\left(\Psi_{x,\la} (\omega ) f(\pi_t (\omega))\right),
\eeq
where $\Psi_\la$ can be computed from Girsanov's formula (see
\cite{LS} p.\ 314) and for $\la$ large enough
reads
\be{Psilaomega}
\Psi_{x,\la} (\omega)
=\exp \left[ \sum_{|y-x|\leq R}
\int_0^t \log\frac{c(y,\omega_s^x)}{c(y,\omega_s)}dN^y_s(\omega)
+\sum_{|y-x|\leq R}\int_0^t [c(y,\omega_s)-c(y,\omega^x_s)]ds\right],
\ee
where $N^y_t (\omega)$ is the number of spin flips at site $y$ up to time $t$
along the trajectory $\omega$. We thus obtain the representation of (\ref{pathspaceRN})
by observing that $\Psi_{x,\la}$ does not depend on $\la$ for $\la$ large enough and using
the convergence of $\pee^\la_\si$ to $\pee_\si$ as $\la\uparrow\Z^d$. Indeed, the only
point to check is that
\be{Psileftfrac}
\left(\frac{d\nu^x}{d\nu}\circ\pi_0\right) \Psi_x \in L^1 (\pee_\nu ),
\ee
so that the conditional expectation in (\ref{pathspaceRN}) is well-defined. However,
this is a consequence of the following two observations:
\begin{enumerate}
\item
$d\nu^x/d\nu $ is uniformly bounded because $\nu\in\gee$.
\item
For $\Psi_x$ we have the bound
\be{ineq}
|\Psi_x (\omega )|\leq e^{2Ct}\left(\frac{M}{\epsilon}\right)^{N^{R,x}_t(\omega)},
\ee
where,
as before, $M$ and $\epsilon$ are the maximum and minimum rates,
$N^{R,x}_t(\omega)$ is the total number of spin flips in the region
$\{y:|y-x|\leq R \}$ up to time $t$ along the trajectory $\omega$. Since the rates
are bounded from above, the expectation of the RHS of (\ref{ineq}) over $\pee_\sigma$
is finite uniformly in $\sigma$.
\end{enumerate}
\epr

\subsection{Backwards process}
\label{S3.3}

Proposition \ref{prop 2} provides us with a representation of the RN-derivative
$d\nu_t^x/d\nu_t$ that can be interpreted as the expectation of a continuous function
on path space {\it in the backwards process}. The backwards process
is the Markov process with a time-dependent transition operator given by
\be{Tnust}
(T_\nu (s,t) f)(\sigma ) = \E_\nu (f\circ\pi_s|\sigma_t=\sigma) \qquad 0\leq s\leq t,
\ee
where $(\cdot|\si_t=\si)$ is conditional expectation with
respect to the $\si$-field at time $t$. Notice that this transition operator depends on the initial Gibbs measure $\nu$ and is a
function of $s$ and $t$ (time-inhomogeneous process).
Although the evolution has a reversible measure $\mu$, at any finite time
the distribution at time $t$ is not $\mu$. 
This causes essential differences between the forward and the backwards process.

The dependence of $T_\nu (s,t)$ on $\nu$ is crucial
and shows that even for innocent dynamics, like the independent spin-flip process,
the transition operators of the backwards process may fail to be Feller for certain
choices of $\nu$ (see Section \ref{S5} below).
In general, the independence of the Poisson clocks that govern where
the spins are flipped (in the backwards process this  means {\em were flipped}) is lost.

In order to have continuity of the
RN-derivative $d\nu_t^x/d\nu_t$, it is sufficient that the operators $T_\nu (s,t)$
have the Feller property, i.e., map continuous functions to continuous functions.

\bp{pimpl}
If $\nu$ is a Gibbs measure, then:
\be{TnustC}
T_\nu (s,t)C(\Omega )\subset C(\Omega) ~\forall ~0 \leq s < t \leq t_0
\qquad \Longrightarrow \qquad \nu S(t)\in\gee ~\forall ~0 \leq t \leq t_0.
\ee
\ep

\bpr
This is an immediate consequence of Proposition \ref{prop 2}. See also \cite{K}.
\epr

As in Section \ref{S3.1}, we can thus hope to approximate the transition operators of the
backwards process by ``local operators" (operators mapping $\loc$ onto $\loc$).

\bp{prop 4}
For any $\si\in \Omega$ and $0\leq s< t$,
\be{Tnustfsigma}
(T_\nu (s,t) f) (\sigma)=\lim_{\la \uparrow\Z^d}
\frac{\E_\sigma \left(\exp[H^{\mu,\nu}_\la (\sigma_t)] f(\sigma_{t-s})\right)}
{\E_\si \left(\exp[H^{\mu,\nu}_\la (\si_t )]\right)},
\ee
where this equality is to be interpreted as follows: if the limit in the RHS of
(\ref{Tnustfsigma}) is a limit in the uniform topology, then it defines a continuous
version of the LHS.
\ep

\bpr
Let us first compute $T_\nu (s,t)$ in the case of the finite-volume reversible
Markov chain with generator (\ref{finvolgen}). For the sake of notational simplicity,
we omit the indices $\la$ referring to the finite volume, and abbreviate
$\nu_s= \nu S(s)$:
\beq{Tnustf}
(T_\nu (s,t) f) (\sigma)
&=& \sum_\eta p_{t-s} (\eta,\sigma) \frac{\nu_s (\eta)}{\nu_t (\sigma)} f(\eta)\nonumber\\
&=& \frac{ \mu_t (\sigma )}{\nu_t (\sigma)}
\sum_\eta p_{t-s} (\sigma,\eta) \frac{\nu_s (\eta)}{\mu_s (\eta) }
f(\eta)\nonumber\\
&=&
\left[S(t)\left(\frac{d\nu}{d\mu}\right) (\si)\right]^{-1}
\sum_{\eta} p_{t-s} (\sigma,\eta) \left[
S(s)\left(\frac{d\nu}{d\mu}\right)(\eta)\right]f(\eta)\nonumber\\
&=&
\frac{ S(t-s)\left( S(s)\left(\frac{d\nu}{d\mu}\right)f\right)}
{S(t)\left( \frac{d\nu}{d\mu}\right)}(\sigma)\nonumber\\
&=&
\frac{\E_\sigma \left(\exp[H^{\mu,\nu}_\la (\sigma_t)] f(\sigma_{t-s} )\right)}
{\E_\si \left(\exp[H^{\mu,\nu}_\la (\si_t)]\right)},
\eeq
where $H^{\mu,\nu}_\la$ is defined in Definition \ref{difham}.
\epr

Propositions \ref{pimpl} and \ref{prop 4} are the analogues of Propositions
\ref{Kozlov} and \ref{prop 1}. We will not actually use them, but they provide
useful insight.

\subsection{Criterion for Gibbsianness of $\nu S(t)$}
\label{S3.4}

A useful tool to study whether $\nu S(t)\in \gee$ is to consider the joint
distribution of $(\sigma_0,\sigma_t)$, where $\sigma_0$ is distributed according
to $\nu$. Let us denote this joint distribution by $\hat{\nu}_t$, which can be
viewed as a distribution on $\{-1, +1\}^S$ with $S=\Z^d \oplus \Z^d$ consisting
of two ``layers" of $\Z^d$. The correspondence between $\hat{\nu}_t$ and $\nu S(t)$
is made explicit by the formula
\be{inthatnut}
\int \hat{\nu_t}(d\sigma,d\eta) f(\sigma)g(\eta)
= \int \nu(d\si ) (f S(t)g)(\sigma) \qquad f,g\in\loc.
\ee
Now, for reasons that will become clear later,
$\hat{\nu_t}$ has more chance of being Gibbsian than $\nu S(t)$. The latter
can then be viewed as the restriction of a Gibbs measure of a two-layer system to the second layer.
Restrictions of Gibbs measures have been studied e.g.\ in \cite{Schon}, \cite{MV94} \cite{FerPfis},
\cite{MRV}, \cite{MRSV}, and it is well-known that they can fail to be Gibbsian,
and most examples of non-Gibbsian measures can be viewed as restrictions of Gibbs measures. Formally, the Hamiltonian of $\hat{\nu_t}$ is
\be{jointham}
H_t(\sigma,\eta)= H_\nu (\sigma)-\log p_t(\sigma,\eta),
\ee
where $p_t(\sigma,\eta)$ is the transition kernel of the dynamics. Of course, the object
$\log p_t(\sigma,\eta)$ has to be interpreted in the sense of the formal sums
$\sum_{A} U(A,\sigma)$ introduced in Section \ref{S2.3}. More precisely,
if $\delta_\sigma S(t)$ is a Gibbs measure for any $\sigma$, then $\log p_t
(\sigma,\eta )$ is the Hamiltonian of this Gibbs measure. In order to prove or
disprove Gibbsianness of the measure $\nu S(t)$, one has to study the Hamiltonian
(\ref{jointham}) for {\it fixed} $\eta$. Let us denote by $\gee (H^t_\eta)$
the set of Gibbs measures associated with the Hamiltonian $H^t_\eta= H_t(\cdot,\eta)$.
From \cite{FerPfis} we have the following:

\bp{phasetrans*}
For any $t\geq 0$:
\begin{enumerate}
\item
If $|\gee (H^t_\eta)|=1$ for all $\eta\in \Omega$, then $\nu S(t)$
is a Gibbs measure.
\item
For monotone specifications, if $|\gee (H^t_\eta)| \geq 2$, then
$\eta$ is a bad configuration for $\nu S(t)$,  so $\nu S(t)$ is not a Gibbs measure
(by Proposition \ref{Kozlov}).
\end{enumerate}
\ep

\bpr
See \cite{FerPfis}. Part 2 is expected to be true without the
requirement of monotonicity but this has not been proved.
\epr

\noindent
A monotone specification arises e.g.\ when the Hamiltonian of (\ref{jointham}) comes
from a ferromagnetic pair potential and an arbitrary single-site part (possibly an
inhomogeneous magnetic field).

In the case of a high-temperature dynamics ($T_\mu >>1$), $\delta_\sigma S(t)$ converges
to $\mu$ for any $\sigma$. This implies that for large $t$ we can view the Hamiltonian
of (\ref{jointham}) as follows:
\be{HtsigmaetaHnusigma}
H_t (\sigma,\eta) = H_\nu (\sigma ) + H_\mu (\eta ) + o_{\sigma,\eta}(t),
\ee
where $o_{\sigma,\eta}(t)$ means some Hamiltonian with corresponding interaction converging
to zero as $t\uparrow\infty$ in $\bee$. Therefore, if $H_\nu$ does not have a phase
transition, then $H^t_\eta$ should not have a phase transition either for large $t$.
On the other hand, if $H_\nu$ does have a phase transition, then the
$o_{\sigma,\eta}(t)$-term will be important to {\it select one of the phases}. In
Sections \ref{S5}--\ref{S6} we will come back to this description in more detail.

The case of independent spin flips corresponds to $H_\mu=0$.


\section{Conservation of Gibbsianness for small times}
\label{S4}

Having put the technical machinery in place in Sections \ref{S2}--\ref{S3},
we are now ready to formulate and prove our main results in Sections
\ref{S4}--\ref{S6}.

In this section we prove that for every finite-range spin-flip dynamics starting from a
Gibbs measure $\nu$ corresponding to a finite-range interaction the measure
$\nu S(t)$ remains Gibbsian in a small interval of time $[0,t_0]$. The intuition
behind this theorem is that for small times the set of sites where a spin flip
has occurred consists of ``small islands'' that are far apart in a ``sea'' of
sites where no spin flip has occurred. This means that sites that are far apart
have more or less disjoint histories.

\bt{Gibbscons}
Let both the initial measure $\nu$ and the reversible measure $\mu$ be Gibbs
measures for finite-range interactions $U_\nu$ resp.\ $U_\mu$. Then there exists
$t_0=t_0 (\mu,\nu)>0$ such that $\nu S(t)$ is a Gibbs measure for all $0\leq t \leq
t_0$.
\et

\bpr
During the proof we abbreviate $H_\la=H^{\mu,\nu}_\la$.
We prove that the limit
\be{thelim}
\lim_{\la\uparrow\Z^d}\frac{\E_{\si^x}\left(\exp[H_\la (\si_t)]\right)}
{\E_{\si} \left(\exp[H_\la (\si_t)]\right)}
\ee
converges uniformly in $t\in[0,t_0]$ for $t_0$ small enough when $U_\nu,U_\mu \in
\bee^{fr}$. The $t_0$ depends on both $U_\nu$ and $U_\mu$.

Let us write $R_\nu,R_\mu$ to denote the range of $U_\nu,U_\mu$ (see Section
\ref{S2.2}).

\medskip\noindent
{\bf I: $R_\nu<\infty$, $R_\mu=0$.}

\medskip\noindent
To warm up, we first deal with unbiased independent spin-flip dynamics. For this
dynamics the distribution of $\si_t$ under $\pee^0_{\si^x}$ coincides with the
distribution of $\si^x_t$ under $\pee^0_\si$. Therefore we can write
\beq{eq1}
\frac{\E^0_{\si}\left(\exp[H_\la (\si^x_t)]\right)}
{\E^0_{\si}\left(\exp[H_\la (\si_t)]\right)}
&=&
\frac{
\sum_{A\subset\la} \delta_t^{|A|} (1-\delta_t)^{|\la|-|A|}
\exp[(H^{A\Delta\{x\}}-H)(\si)]}
{\sum_{A\subset\la} \delta_t^{|A|} (1-\delta_t)^{|\la|-|A|}
\exp([H^{A}-H)(\si)]}\nonumber\\
&=&
\left(\frac{
\sum_{A\subset\la} \left(\frac{\delta_t}{1-\delta_t}\right)^{|A|}
\exp[(H^{A\Delta\{x\}}-H^{\{x\}})(\si)]}
{\sum_{A\subset\la} \left(\frac{\delta_t}{1-\delta_t}\right)^{|A|}
\exp[(H^{A}-H)(\si)]}\right)\Psi_x (\si),
\eeq
where
\be{Psixsi}
\Psi_x (\si ) = \exp[(H^{\{x\}}-H)(\si )]
\ee
is a continuous function of $\si$, the sum runs over
\be{Adef}
A = \{y\in\Lambda : \si_t (y) \not=\si_0 (y)\},
\ee
while
\be{deltatdef}
\delta_t = \pee^0_\si (\si_t (x)\not=\si_0(x)) = 1-e^{-2t}.
\ee
The notation $H^A$, $A\subset\la$, is defined by
\be{HAsi}
H^A(\si) = H(\si^A)
\ee
with $\si^A$ the configuration obtained from $\si$ by flipping all the spins in $A$.

Suppose first that $R_\nu=1$. Then
\be{HAB}
H^{A\cup B}- H^A = H^B - H \qquad \forall~A,B:~d(A,B)>1.
\ee
For $A\subset \la$ we can decompose $A$ into disjoint nearest-neighbor connected subsets
$\gamma_1,\ldots,\gamma_k$ and thus rewrite (\ref{eq1}) as follows:
\be{eq2}
\frac{\E^0_{\si} \left(\exp[H_\la (\si^x_t)]\right)}
{\E^0_{\si} \left(\exp[H_\la (\si_t)]\right)}
= \left(
\frac{
\sum_{n=0}^\infty
\frac{1}{n!}
\sum_{\gamma_1,\ldots,\gamma_n\subset\la,\gamma_i\cap\gamma_j=\emptyset}
\prod_{i=1}^n w_\si^x(\gamma_i)}
{\sum_{n=0}^\infty
\frac{1}{n!}
\sum_{\gamma_1,\ldots,\gamma_n\subset\la,\gamma_i\cap\gamma_j=\emptyset}
\prod_{i=1}^n w_\si(\gamma_i)
}
\right)
\Psi_x
\ee
with
\be{wsi}
\begin{array}{lll}
w_\si^x (\gamma) &=& \epsilon_t^{|\gamma|}
\exp[H^{\gamma\Delta\{x\}}(\si ) - H^{\{x\}}(\si )]\\
w_\si (\gamma ) &=& \epsilon_t^{|\gamma|}\exp [H^\gamma (\si)-H (\si)]
\end{array}
\ee
and $\epsilon_t = \delta_t /(1-\delta_t)$. Note that $w_\si^x (\gamma)= w_\si (\gamma)$
for all $\gamma $ that do not contain $x$.

Next, since
\be{HgammaH}
|(H^\gamma -H) (\si )|\leq |\gamma|C
\ee
with
\be{C2}
C=2\sup_\la\sup_\si \frac{|H_\la (\si)|}{|\la|} <\infty,
\ee
we have the estimate
\be{estim}
|w_\si (\gamma)|\leq \exp(-\alpha_t |\gamma|) \qquad \mbox{ with }
\alpha_t = -C + \log(1/\epsilon_t).
\ee
A similar estimate holds for $|w_\si^x (\gamma)|$. Since $\alpha_t\uparrow\infty$ as
$t\downarrow 0$, it follows that for $t$ small enough we can expand the logarithm of
both the numerator and the denominator in (\ref{eq2}) in a uniformly convergent cluster
expansion:
\beq{logsum}
\log \left(\sum_{n=0}^\infty\frac{1}{n!}
\sum_{\gamma_1,\ldots,\gamma_n\subset\la,\gamma_i\cap\gamma_j=\emptyset}
\prod_{i=1}^n w_\si^x(\gamma_i)\right)
&=&
\sum_{\Gamma} a(\Gamma ) w_\si^x (\Gamma ),\nonumber\\
\log
\left(\sum_{n=0}^\infty
\frac{1}{n!}
\sum_{\gamma_1,\ldots,\gamma_n\subset\la,\gamma_i\cap\gamma_j=\emptyset}
\prod_{i=1}^n w_\si(\gamma_i)\right)
&=&
\sum_{\Gamma} a(\Gamma) w_\si (\Gamma).
\eeq
By the estimate (\ref{estim}) we have, for $t$ small enough,
\be{Gammani}
\limsup_{\la\uparrow\Z^d} \sum_{\Gamma\ni x, \Gamma\not\subset\la}
\sup_\si| a(\Gamma )(w^x_\si (\Gamma ) - w_\si (\Gamma )) | = 0
\qquad \forall x\in\Z^d
\ee
and hence we obtain uniform convergence of the limit in (\ref{thelim}).

The case $R_\nu<\infty$ is treated in the same way. We only have to redefine the
$\gamma_i$'s as the $R_\nu$-connected decomposition of $A$. Note that $t_0$ depends
on $R_\nu$ and converges to zero when $R_\nu\uparrow\infty$.

\medskip\noindent
{\bf II: $R_\nu<\infty$, $R_\mu<\infty$.}

\medskip\noindent
Next we prove that the limit (\ref{thelim}) converges uniformly if both interactions
$U_\mu,U_\nu$ are finite range. For the sake of notational simplicity we first restrict
ourselves to the case $R_\nu=R_\mu=1$.

We abbreviate $U= U_\mu - U_\nu$. The idea is that we go back to the independent
spin-flip dynamics via Girsanov's formula. After that we can again
set up a cluster expansion,
which includes additional factors in the weights due to the dynamics.

The first step is to rewrite (\ref{thelim}) in terms of the independent spin-flip
dynamics:
\beq{Girsa}
&&\frac{\E_{\si^x}\left(\exp[H_\la (\si_t)]\right)}
{\E_{\si}\left(\exp[H_\la (\si_t)]\right)}\\
&&\qquad =
\frac{\E^0_\si\left(
\exp \left(\sum_{y\in \la}\int_0^t \log c(y,\si_s^x) dN^y_s +
\int_0^t (1-c(y,\si_s^x)) ds
\right)
\exp[H_\la (\si_t^x )]\right)}
{\E^0_\si\left(
\exp \left(\sum_{y\in \la}\int_0^t \log c(y,\si_s) dN^y_s +
\int_0^t (1-c(y,\si_s)) ds
\right)
\exp[H_\la (\si_t )]\right)}.\nonumber
\eeq
For a given realization $\omega$ of the independent spin-flip process, we say that a site
$y$ is {\it $\omega$-active} if the spin at that site has flipped at least once. The set
of all $\omega$-active sites is denoted by $J(\omega)$. Let $\bar{\sigma}$ denote the
trajectory that stays fixed at $\sigma$ over the time interval $[0,t]$.  For $A\subset \la$,
define
\be{stpot1}
\begin{array}{llll}
\uu_1 (A,\omega)
&=& \int_0^t \log c(y,\omega_s ) dN^y_s (\omega)+\int_0^t (1-c(y,\omega_s))ds
&\mbox{if } A = D_{c_y}\\
&=& 0
&\mbox{if } A \not= D_{c_y},\\
&&&\\
\uu_2 (A,\omega) &=& U(A,\omega_t), &
\end{array}
\ee
and put
\be{stpot}
\uu (A,\omega) = \uu_1 (A,\omega) + \uu_2 (A,\omega).
\ee
Also define
\be{flipstpot}
\uu^x (A,\omega ) = \uu (A,\omega^x),
\ee
where the trajectory $\omega^x$ is defined as
\be{omegax}
(\omega^x)_s = (\omega_s)^x \qquad 0\leq s\leq t.
\ee
With this notation we can rewrite the right-hand side of (\ref{Girsa}) as
\be{tussen}
\left(
\frac{\E^0_\si \left(\exp \left(
\sum_{A\subset\la} [\uu^x (A,\omega) -\uu^x (A,\bar{\si} )]\right)\right)}
{\E^0_\si \left(\exp \left(
\sum_{A\subset\la} [\uu (A,\omega) -\uu (A,\bar{\si} )]\right)\right)}
\right)
\Psi_x (\sigma ),
\ee
where
\be{Psixsinog}
\Psi_x (\si ) = \exp \left(\sum_{A\ni x} [\uu(A,\bar{\si})-\uu (A,\bar{\si}^x)]\right)
\ee
is a continuous function of $\si$. In order to obtain the uniform convergence of
(\ref{thelim}), it suffices now to prove the uniform convergence of the expression
between brackets in (\ref{tussen}).

As in part I, we decompose the set of $\omega$-active sites into disjoint
nearest-neighbor connected sets $\gamma_1,\ldots,\gamma_k$ and rewrite, using the product character of $\E^0_\si$,
\beq{fracE0}
&&\frac{\E^0_\si \left(\exp \left(
\sum_{A\subset\la} [\uu^x (A,\omega) -\uu^x (A,\bar{\si} )]\right)\right)}
{\E^0_\si \left(\exp \left(
\sum_{A\subset\la} [\uu (A,\omega) -\uu (A,\bar{\si} )]\right)\right)}
\nonumber\\
&&\qquad\qquad =
\frac{\sum_{n=0}^\infty
\frac{1}{n!}
\sum_{\gamma_1,\ldots,\gamma_n\subset\la,\gamma_i\cap\gamma_j=\emptyset}
\prod_{i=1}^n w_\si^x(\gamma_i) }
{\sum_{n=0}^\infty
\frac{1}{n!}
\sum_{\gamma_1,\ldots,\gamma_n\subset\la,\gamma_i\cap\gamma_j=\emptyset}
\prod_{i=1}^n w_\si(\gamma_i) }.
\eeq
The cluster weights are now given by
\be{neww}
w_\si (\gamma ) = e^{t|\gamma|}\E^0_\si\left( 1\{J(\omega)\supset\gamma\}
\exp \left(\sum_{A\cap \gamma \not=\emptyset}
[\uu(A,\omega_\gamma\bar{\si}_{\la\setminus\gamma})- \uu(A,\bar{\si})]
\right)\right),
\ee
and an analogous expression for $w^x_\si$ after we replace $\uu$ by $\uu^x$.
The factor $e^{t|\gamma|}$ arises from the probability
\be{E0si}
\pee^0_\si \left(J(\omega)^c \supset \la\setminus\cup_{i}\gamma_i\right)
= e^{-t|\la\setminus\cup_{i}\gamma_i|}
= e^{-t|\Lambda|} \prod_i e^{t|\gamma_i|}.
\ee
Having arrived at this point, we can proceed as in the case of the independent spin-flip
dynamics. Namely, we estimate the weights $w_\si$ and prove that
\be{cruxesti}
w_\si (\gamma ) \leq e^{-\alpha_t |\gamma|}
\ee
with $\alpha_t\uparrow\infty$ as $t\downarrow 0$. To obtain this estimate, note that
\be{pee0}
\pee^0_\si (J(\omega)\supset\gamma)\leq (1-e^{-t})^{|\gamma|}.
\ee
Then apply to (\ref{neww}) Cauchy-Schwarz, the bounds $\epsilon \leq c_y\leq M$ on the flip
rates, and the estimate
\be{Cest}
C=\sup_{\la}\sup_\si \frac{1}{|\la|}\sum_{A\cap \la\not=\emptyset} |U(A,\si)|
<\infty,
\ee
to obtain
\be{wsigamma}
w_\si (\gamma ) \leq e^{Kt|\gamma|} (1-e^{-t})^{-\frac{1}{2}|\gamma|}
\qquad \mbox{ for some } K=K(\epsilon,M,C).
\ee
This clearly implies (\ref{cruxesti}).

The case $R_\nu,R_\mu<\infty$ is
straightforward
after redefining the $\gamma_i$'s.
\epr


\section{Infinite-temperature dynamics}
\label{S5}

\subsection{Set-up}
\label{S5.1}

In this section we consider the evolution of a Gibbs measure $\nu$
under a product dynamics, i.e., the flip rates $c(x,\sigma)$ depend only
on $\sigma (x)$. The associated process $\{\sigma_t: t\geq 0\}$ is
a product of independent Markov chains on $\{-1,+1\}$:
\be{peesigma}
\pee_\sigma = \otimes_{x\in\Z^d} \pee_{\sigma (x)},
\ee
where $\pee_{\sigma (x)}$ is the Markov chain on $\{-1,+1\}$ with generator
\be{Lxvi}
L_x \vi (\alpha ) = c(x,\alpha ) [\vi (-\alpha ) - \vi (\alpha)].
\ee
Let us denote by $p^x_t (\alpha,\beta )$ the probability for this Markov chain
to go from $\alpha$ to $\beta$ in time $t$. The Hamiltonian (\ref{jointham}) of the
joint distribution of $(\si_0,\si_t)$ is then given by
\be{Htsigmaeta}
H_t (\sigma, \eta )= H_\nu (\sigma ) - \sum_x \log p^x_t (\sigma (x),\eta (x)).
\ee
This equation can be rewritten as
\be{prodham}
H_t (\si,\eta )= H_\nu (\si ) - \sum_x h^x_1 (t)\si (x)
- \sum_x h^x_2 (t) \eta (x) - \sum_x h^x_{12} (t) \si (x) \eta (x)
\ee
with
\beq{thefields}
h^x_1 (t) &=&
\frac{1}{4} \log \frac{p^x_t (+,+) p^x_t (+,-)}{p^x_t (-,+) p^x_t (-,-)}
\nonumber\\
h^x_2 (t) &=&
\frac{1}{4} \log \frac{p^x_t (+,+) p^x_t (-,+)}{p^x_t (+,-) p^x_t (-,-)}
\nonumber\\
h^x_{12} (t) &=&
\frac{1}{4} \log \frac{p^x_t (+,+) p^x_t (-,-)}{p^x_t (+,-) p^x_t (-,+)}.
\nonumber\\
\eeq
The fields $h^x_1$ resp.\ $h^x_2$ tend to pull $\sigma$ resp.\ $\eta$ in their direction,
while $h^x_{12}$ is a {\it coupling} between $\sigma$ and $\eta$ that tends to align
them. Indeed, note that $h^x_{12} (t)$ is positive because
\be{pxt}
p^x_t (+,+) p_t^x(-,-) - p^x_t(+,-) p^x_t (-,+)
= \mbox{det}(\exp (tL_x))\geq 0.
\ee
In what follows we will consider the case where the single-site generators $L_x$ are
independent of $x$ and are given by
\be{Lxfrac}
L =\frac{1}{2}
\left(\begin{array}{cc}
-1+\epsilon & 1-\epsilon\\
1+\epsilon & -1-\epsilon
\end{array} \right)
\qquad \mbox{ for some } 0 \leq \epsilon < 1.
\ee
For $\epsilon>0$ this means independent spin flips favoring plus spins, for $\epsilon=0$
it means independent unbiased spin flips. The invariant measure of the single-site Markov
chain is $(\nu(+),\nu(-))=\frac{1}{2}(1+\epsilon,1-\epsilon)$. The relevant parameter in
what follows is
\be{delta*}
\delta= \frac{\nu(-)}{\nu(+)}=\frac{1-\epsilon}{1+\epsilon}.
\ee
In terms of this parameter the fields in (\ref{thefields}) become
\beq{deltafields}
h_1(t)&=&\frac{1}{4}\log\left(\frac{1+\delta e^{-t}}{1+\frac{1}{\delta}
e^{-t}}\right)\nonumber\\
h_2 (t)&=&-\frac{1}{2}\log\delta + h_1 (t)\nonumber\\
h_{12} (t)&=& \frac{1}{4}\log
\frac{(1+\delta e^{-t})(1+\frac{1}{\delta} e^{-t})}{(1-e^{-t})^2}.
\eeq
In particular, for $\delta =1$ we get $h_1(t)=h_2(t)=0$ and
\be{Kadfield}
h_{12} (t)=\frac{1}{2}\log \frac{1+e^{-t}}{1-e^{-t}}.
\ee

\subsection{ $1<<T_\nu \leq \infty$, $T_\mu =\infty$}
\label{S5.2}

\bt{Dobr}
Let $\nu$ be a high- or infinite-temperature Gibbs measure, i.e., its interaction
$U_\nu$ satisfies (\ref{DobU}). Let $S(t)$ be the semigroup of an arbitrary
infinite-temperature dynamics. Then $\nu S(t)$ is a Gibbs measure for all $t\geq 0$.
\et

\bpr
The joint distribution of $(\si_0,\si_t)$ is Gibbs with Hamiltonian (recall (\ref{jointham})
and (\ref{prodham}))
\be{Htsigmaeta2}
H_t (\sigma,\eta)= H_\nu (\sigma) + \sum_x [h_1(t)+h_{12}(t)\eta(x)]\sigma(x)
+ h_2(t) \sum_x \eta_x.
\ee
For fixed $\eta$, the last term is constant in $\sigma$ and can therefore be forgotten.
Since $H_t (\cdot,\eta)$ differs from $H_\nu(\cdot)$ only in the single-site interaction,
$H_t (\cdot,\eta)$ satisfies (\ref{DobU}) if and only if $H_\nu(\cdot)$ satisfies
(\ref{DobU}). Hence $|\gee (H_t (\cdot,\eta)|=1$ for any $\eta$, and we conclude from
Proposition \ref{phasetrans*} that $\nu S(t)$ is Gibbsian.
\epr

Theorem \ref{Dobr} should not come as a surprise: the infinite-temperature dynamics act
as a single-site Kadanoff transformation and in the Dobrushin uniqueness regime such
renormalized measures stay Gibbsian
\cite{GP}, \cite{isr}, \cite{Enter}.

\subsection{$0<T_\nu<<1$, $T_\mu=\infty$, $\delta=1$}
\label{S5.3}

For the initial measure we choose the low-temperature plus-phase of the
$d$-dimensional Ising model, $\nu=\nu_{\beta,h}$, i.e., the Hamiltonian $H_\nu$
is specified to be
\be{Isingham}
H_\nu (\sigma ) = -\beta\sum_{<x,y>} \sigma (x)\si (y) - h\sum_x \si (x),
\ee
where $\sum_{<x,y>}$ denotes the sum over nearest-neighbor pairs, and $\beta>>\beta_c$
with $\beta_c$ the critical inverse temperature. The dynamics has generator
\be{Lfsumx}
Lf= \sum_x \nabla_x f,
\ee
corresponding to the case $\delta=1$. The joint measure has Hamiltonian as in
(\ref{Htsigmaeta2}), with $h_1(t)=h_2(t)=0$ and $h_{12}(t)=h_t$:
\be{jointIham}
H_t (\si,\eta) = -\beta\sum_{<x,y>} \si (x) \si (y) - h\sum_x \si( x)
-h_t\sum_x \si (x)\eta (x).
\ee
The ``dynamical field'' is given by $h_t=-(1/2)\log[\tanh (t/2)]$.

\bt{Isingthm}
For $\beta >>\beta_c$:
\begin{enumerate}
\item
There exists a $t_0=t_0(\beta,h)$ such that $\nu_{\beta,h} S(t)$ is a
Gibbs measure for all $0 \leq t\leq t_0$.
\item
If $h>0$, then there exists a $t_1= t_1 (\beta,h)$ such that $\nu_{\beta,h} S(t)$
is a Gibbs measure for all $t\geq t_1$.
\item
If $h=0$, then there exists a $t_2 = t_2 (\beta)$ such that $\nu_{\beta,0} S(t)$
is not a Gibbs measure for all $t\geq t_2$.
\item
For $d\geq 3$, if $h\leq h(\beta)$ small enough, then there exist $t_3= t_3 (\beta,h)$ and $t_4 = t_4 (\beta, h)$ such
that $\nu_{\beta,h} S(t)$ is not a Gibbs measure for all $t_3 \leq t \leq t_4$.
\end{enumerate}
\et

\bpr
The proof uses (\ref{jointIham}).

\medskip\noindent
1.\ For small $t$ the dynamical field $h_t$ is large and, for {\it given} $\eta$,
forces $\sigma$ in the direction of $\eta$. Rewrite the joint Hamiltonian in
(\ref{jointIham}) as
\beq{Htsietasqrt}
H_t (\si,\eta ) &=& \sqrt{h_t}\left(
- \frac{\beta}{\sqrt{h_t}}\sum_{<x,y>}\si (x)\si (y)
- \frac{h}{\sqrt{h_t}}\sum_{x} \si (x)
- \sqrt{h_t}\sum_x\si (x)\eta (x)\right)
\nonumber\\
&=& \sqrt{h_t} \tilde{H_t} (\sigma,\eta).
\eeq
For $0\leq t\leq t_0$ small enough, $\tilde{H_t}$ has the unique ground state $\eta$
and so, for $\lambda\geq\lambda_0$ large enough, $\lambda\tilde{H_t}$ satisfies
(\ref{DobU}) (see \cite{Geo}, example 2, p.\ 147). Therefore, for $0\leq t\leq t_1$
such that $\sqrt{h_t}\geq\lambda_0$, $H_t(\cdot,\eta)$ has a unique Gibbs measure
for any $\eta$. Hence, $\nu S(t)$ is Gibbs by Proposition \ref{phasetrans*}(1).

\medskip\noindent
2.\ For large $t$ the dynamical field $h_t$ is small and cannot cancel the
effect of the external field $h>0$. Rewrite the joint Hamiltonian as
\beq{Htsietasqrtbeta}
H_t (\si,\eta ) &=&
\sqrt{\beta} \left(
- \sqrt{\beta}\sum_{<x,y>} \si (x)\si (y)
- \frac{h}{\sqrt{\beta}} \sum_x \si (x)
- \frac{h_t}{\sqrt{\beta}}\sum_x \si (x) \eta (x)\right)
\nonumber\\
&=&
\sqrt{\beta}\tilde{H_t} (\si,\eta ).
\eeq
For $t\geq t_1$ large enough (independently of $\beta$), $\tilde{H_t}(\cdot,\eta)$ has the
unique ground state $\si =h/|h|$. Hence, for $\beta$ large enough, $\sqrt{\beta}\tilde{H_t}
(\cdot,\eta )$ has a unique Gibbs measure by (\ref{DobU}) (again, see \cite{Geo}, example 2,
p.\ 147). Hence, $\nu S(t)$ is Gibbs by Proposition \ref{phasetrans*}(1).

\medskip\noindent
3.\ This fact is a consequence of the results in \cite{Enter}, section 4.3.4,
for the single-site Kadanoff
transformation. Since the joint Hamiltonian in (\ref{jointIham}) is ferromagnetic, it
suffices to show that there is a special configuration $\eta_{spec}$ such that
$|\gee (H(\cdot,\eta_{spec})|\geq 2$. We choose $\eta_{spec}$ to be the alternating
configuration. For $t\geq t_2$ large enough, $H_t (\cdot,\eta_{spec})$ has two ground states,
and by an application of Pirogov-Sinai theory (see \cite{Enter} Appendix B), it follows
that, for $\beta$ large enough, $|\gee (H_t(\cdot,\eta_{spec})|\geq 2$. Therefore
$\eta_{spec}$ is a bad configuration for $\nu S(t)$, implying that $\nu S(t)$ is not Gibbs
by Proposition \ref{phasetrans*}(2).

\medskip\noindent
4.\ In this case we rewrite the Hamiltonian in (\ref{jointIham}) as
\be{Htsietabetasum}
H_t (\si,\eta)=-\beta\sum_{<x,y>}\si (x)\si (y)
-\sum_{x}[h+h_t\eta(x)]\si (x).
\ee
For ``intermediate" $t$ we have that $h$ and $h_t$ are of the same order. As explained in
\cite{Enter} section 4.3.6, we can find a bad configuration $\eta_{spec}$ such that the term $\sum_x h_t \eta(x)\si (x)$ in the Hamiltonian
``compensates" the effect of the homogeneous-field term $\sum_x h\si (x)$ and for which $H_t
(\cdot,\eta_{spec} )$ has
two ground states which are predominatly plus and minus.
Since the proof of existence of $\eta_{spec}$ requires analysis of
the random field Ising model, we have to restrict to the case $d\geq 3$
(unlike the previous case $\eta_{spec}$ is not constructed,
but chosen from a measure one set). Then for $\beta$ large enough, by a
Pirogov-Sinai
argument (see appendix B, Theorem B 31 of \cite{Enter})
$|\gee (H_t(\cdot,\eta_{spec})|\geq 2$, implying that $\nu S(t)$ is not Gibbs by
Proposition \ref{phasetrans*}(2).
\epr
\begin{flushleft}
{\bf Remark:}
\end{flushleft}
From the estimate (B89) in \cite{Enter}, Appendix B, we
can conclude the following:
\begin{enumerate}
\item $t_0 (\beta,h)\to 0$ as $\beta\to\infty$, and
$t_0(\beta, h)\to\infty$ as $h\to\infty$.
\item $t_2 (\beta)\to 0$ as $\beta\to\infty$.
\item $t_3 (\beta,h)\to 0$ as $\beta\to\infty$.
\end{enumerate}

\subsection{$0<T_\nu<<1$, $T_\mu =\infty$, $\delta<1$}
\label{S5.4}

Let us now consider a biased dynamics. At first sight one might expect this case
to be analogous to the case of an unbiased dynamics with an initial measure
having $h>0$. However, this intuition is false.

\bt{biasdyn}
The same results as in Theorem \ref{Isingthm} hold, but with the $t_i$'s also depending
on $\delta$. For item 4 we need the restrictions $d\geq 3$ and $|h+\frac14 \log \delta|$ small enough.
\et

\bpr
The last term in (\ref{prodham}) being irrelevant, we can drop it and study the
Hamiltonian
\be{effham}
\hat{H_t} (\sigma,\eta )
= -\beta\sum_{<x,y>} \si (x) \si (y) -
\sum_x \si(x) \left[(h+h_1(t)) + h_{12}(t)\eta (x)\right].
\ee
This Hamiltonian is of the same form as (\ref{jointIham}), but with $h$ becoming
$t$-dependent. We have $\lim_{t\uparrow\infty}h_1(t)=0$ and
$\lim_{t\uparrow\infty}h_{12}(t)=0$ with
\be{frach12}
\lim_{t\uparrow\infty} \frac{h_{12} (t)}{h_1(t)} = \frac{1+\delta}{1-\delta}>1,
\ee
so that, in the regime where $\beta>>\beta_c$, $h=0$, $t>>1$, we find that
the effect of $h_{12}(t)$ dominates. Hence we can find a special
configuration that compensates the effect of the field $h_1 (t)$ and for which the Hamiltonian (\ref{effham}) has two ground states, implying that $\nu S(t)\not\in \gee$. Similarly, when $h>0$ we can find $t$ intermediate
such that $\sum_x (h+h_1(t)) \si (x)$ is ``compensated" by $\sum h_{12} (t)\si (x) \eta (x)$.
\epr
\begin{flushleft}
{\bf Remark:}
\end{flushleft}
Note that if $T_\nu=0$, $T_\mu=\infty$, then $\nu S(t)$ is a product measure
for all $t>0$ and hence is Gibbs.


\section{High-temperature dynamics}
\label{S6}

\subsection{Set-up}

In this section we generalize our results in Section \ref{S5} for the infinite-temperature
dynamics to the case of a high-temperature dynamics. The key technical tool is a cluster
expansion that allows us to obtain Gibbsianness of the joint distribution of $(\si_0,\si_t)$
with a Hamiltonian of the form (\ref{jointham}). The main difficulty is to give meaning
to the term $\log p_t(\si,\eta)$, i.e., to obtain Gibbsianness of the measure
$\delta_\si S(t)$ for any $\si$. In the whole of this section we will assume that
the rates $c(x, \si)$ satisfy the conditions in Section \ref{S2.2} and, in addition,
\be{cond1}
c(x,\si ) = 1+\epsilon (x,\si)
\ee
with
\be{cond2}
\begin{array}{ll}
&\sup_{\si,x}|\epsilon (x,\sigma)|=\delta <<1\\
&\epsilon (x,\si ) = \epsilon (x,-\si).
\end{array}
\ee
The latter corresponds to a high-temperature unbiased dynamics, i.e., a small unbiased
perturbation of the unbiased independent spin-flip process. For the initial measure
we consider two cases:
\begin{enumerate}
\item
A high- or infinite-temperature Gibbs measure $\nu$. In that case we will find that
$\nu S(t)$ stays Gibbsian for all $t>0$.
\item
The plus-phase of the low-non-zero-temperature $d$-dimensional Ising model,
$\nu_{\beta,h}$, corresponding to the Hamiltonian in (\ref{Isingham}). In that
case we will find the same transitions as for the infinite-temperature dynamics.
\end{enumerate}

\subsection{Representation of the joint Hamiltonian}
\label{S6.2}

In this section we formulate the main result of the space-time cluster expansion in
\cite{MV93} and \cite{MN}. We indicate the line of proof of this result, and refer
the reader to \cite{MN} for the complete details.

\bt{clust}
Let $\nu$ be a Gibbs measure with Hamiltonian $H_\nu$, and let the dynamics be governed
by rates satisfying (\ref{cond1}--\ref{cond2}). Then the joint distribution of
$(\si_0,\si_t)$, when $\si_0$ is distributed according to $\nu$, is a Gibbs measure
with Hamiltonian
\be{highTham}
H_t (\si,\eta) = H_\nu (\si ) + H^t_{dyn} (\si,\eta).
\ee
The Hamiltonian $H_{dyn}^t (\si,\eta)$ corresponds to an interaction $U^t_{dyn}
(A,\si,\eta)$, $A\in\s$, that has the following properties:
\begin{enumerate}
\item
The interaction splits into two terms
\be{splitpot}
U_{dyn}^t = U_0^t + U^t_{\delta},
\ee
where $U^t_0$ is the single-site potential corresponding to the Kadanoff
transformation:
\be{Ut0}
\begin{array}{llll}
U^t_0 (\{x\},\si,\eta) &=& -\frac{1}{2}\log [\tanh(t/2)]\si (x)\eta (x)
&x\in\Z^d,\\
U^t_0 (A,\si,\eta) &=& 0 \qquad &\mbox{if}\ |A|\not= 1.
\end{array}
\ee
\item
The term $U_\delta^t =U_\delta^t (A,\si,\eta)$ decays exponentially in the
diameter of $A$, i.e., there exists $\alpha (\delta)>0$ such that
\be{convpot}
\sup_{t\geq 0}\sup_x\sum_{A\ni x} \sup_{\si,\eta}e^{\alpha(\delta){\rm diam}(A)}
|U^t_\delta (A,\si,\eta)| <\infty.
\ee
and $\alpha(\delta ) \uparrow\infty$ as $\delta\downarrow 0$.
\item The potential $U^t_{dyn}$ converges exponentially fast to the
potential $U_\mu$ of the high-temperature reversible Gibbs measure:
\be{covpot}
\lim_{t\uparrow \infty}\sup_x\sum_{A\ni x} \sup_{\si,\eta}e^{\alpha(\delta){\rm diam}(A)}
|U^t_\delta (A,\si,\eta)- U_\mu (A,\eta)|=0.
\ee
\item
The term $U_\delta^t$ is a perturbation of the term $U_0^t$, i.e.,
\be{delta}
\lim_{\delta\downarrow 0}\sup_{t\geq 0}\frac{\sup_x\sum_{A\ni x} \sup_{\si,\si',\eta}
|U^t_\delta (A,\si,\eta)-U^t_\delta (A,\si',\eta)|}
{\log[\tanh (t/2)]}=0.
\ee
\end{enumerate}
\et

\medskip\noindent
{\bf Remarks:}
\begin{enumerate}
\item
Equation (\ref{Ut0}) corresponds to the infinite-temperature dynamics (i.e.,
$c\equiv 1$).

\item
Equation (\ref{delta}) expresses that the potential as a function of the
rates $c$ is continuous at the point $c\equiv1$,
and that the Kadanoff term is dominant for $\delta<<1$.
\end{enumerate}

\medskip\noindent
{\bf Main steps in the proof of Theorem \ref{clust} in \cite{MN}:}
\begin{itemize}
\item
{\bf Discretization:} The semigroup $S(t)$ can be approximated in a strong sense
by discrete-time probabilistic cellular automata with transition operators of the form
$P_n(\si'|\si)= \prod_x P_n(\si'(x)|\si)$, where
\be{Pnsi}
P_n(\si'(x)|\si ) = \left[1-\frac{1}{n} c(x,\si)\right]\delta_{\si'(x),\si (x)}
+\frac{1}{n} c(x,\si ) \delta_{\si' (x), -\si (x)}.
\ee
\item
{\bf Space-time cluster expansion for fixed discretization $n$:}
For $n$ fixed the quantity
\be{Psinxsietalog}
\Psi_n^x (\si,\eta )=\log\frac{(d\delta_\si
P_n^{\lfloor nt\rfloor})^x}{(d\delta_\si P_n^{\lfloor nt\rfloor})}
\ee
is defined by the convergent cluster expansion
\be{Psinxsietasum}
\Psi_n^x (\si,\eta)=
\sum_{\Gamma\ni x, \Gamma\in \ce} w^{x,n}_{\si,\eta} (\Gamma ),
\ee
where $\ce$ is an appropriate set of clusters on $\Z^{d+1}$.
\item
{\bf Uniformity in the discretization $n$:}
The functions $\Psi_n^x$ converge uniformly as $n\uparrow\infty$ to a continuous
function $\Psi^x$ (which defines a continuous version of $d\mu^x/d\mu$). This is
shown in two steps:
\begin{enumerate}
\item
{\bf Uniform boundedness:}
\be{ub}
\sup_n\sup_x\sup_{\si,\eta}|\Psi^x_n (\si,\eta)| <\infty.
\ee
\item
{\bf Uniform continuity:}
\be{uc}
\lim_{\la\uparrow\Z^d}\sup_{\zeta,\xi}\sup_x
\sup_n|\Psi_n^x (\si_\la\zeta_{\la^c},\eta_\la\xi_{\la^c} )-
\Psi_n^x (\si,\eta ) | = 0 \qquad \forall \sigma,\eta \in \Omega.
\ee
\end{enumerate}
Equations (\ref{ub}) and (\ref{uc}) imply that $\Psi_n^x$ as a function of $n$
contains a uniformly convergent subsequence. The limiting $\Psi^x$ is independent
of the subsequence, since it is a continuous version of $d\mu^x/d\mu$.
\end{itemize}

\subsection{$1<<T_\nu\leq\infty$, $1<<T_\mu<\infty$}
\label{6.3}

Given the result of Theorem \ref{clust}, the case of a high- or infinite-temperature
initial measure is dealt with via Dobrushin's uniqueness criterion (recall Theorem
\ref{Dobr} in Section \ref{S5.2}).

\bt{Dobr*}
Let $\nu$ be a high-temperature Gibbs measure, i.e., its interaction $U_\nu$ satisfies
(\ref{DobU}). Let the rates satisfy (\ref{cond1}--\ref{cond2}). Then, for $\delta$ small
enough, $\nu S(t)$ is a Gibbs measure for all $t\geq 0$.
\et

\bpr
For fixed $\eta$, the Hamiltonian $H_t(\cdot,\eta)$ of (\ref{highTham}) corresponds
to an interaction $U^{\eta,\delta}_t $. By (\ref{convpot}) and (\ref{delta}), this
interaction satisfies
\beq{suptsupx}
&&\lim_{\delta\downarrow 0}\sup_t \sup_x\sum_{A\ni x} (|A|-1)
\sup_{\si,\si'} |U^{\eta,\delta}_t (\si)-U^{\eta,\delta}_t (\si')|
\nonumber\\
&&\qquad = \sum_{A\ni x} (|A|-1)|U_\nu (\si) - U_\nu (\si')| <2.
\eeq
Therefore, for $\delta$ small enough, (\ref{DobU}) is satisfied for the interaction
$U^{\eta,\delta}_t$ for all $t\geq 0$ and all $\eta$. Hence $|\gee (H_t (\cdot,\eta))|=1$,
and we conclude from Proposition \ref{phasetrans*}(1) that $\nu S(t)\in\gee$.
\epr

\subsection{$0<T_\nu <<1$, $1<<T_\mu<\infty$.}
\label{S6.4}

We consider as the initial measure the plus-phase of the low-temperature Ising model
$\nu_{\beta,h}$, introduced in Section \ref{S5.3}. The joint distribution
of $(\si_0,\si_t)$ has the Hamiltonian
\be{pirham}
H_t (\si,\eta ) = -\beta\sum_{<x,y>} \si (x)\si (y)-h\sum_x \si (x)
-\frac{1}{2}\log[\tanh(t/2)]\sum_x \si (x)\eta(x)+H^\delta_t (\si,\eta),
\ee
where $H^\delta_t$ corresponds to the interaction $U^\delta_{dyn}$ introduced
in (\ref{splitpot}).
The following is the analogue of Theorem \ref{Isingthm}
\bt{Isingthm*}
For $\beta>>\beta_c$ and $0<\delta<<1$:
\begin{enumerate}
\item There exists $t_0= t_0 (\beta,h,\delta)$ such that 
$\nu_{\beta,h} S(t)$ is a Gibbs measure for all $0\leq t\leq t_0$.
\item If $h>0$, then there exists $t_1= t_1 (\beta,h,\delta)$ such that
$\nu_{\beta,h} S(t)$ is a Gibbs measure for all $t\geq t_1$.
\item If $h=0$, then there exists $t_2 = t_2 (\beta,\delta)$ such that
$\nu_{\beta,0} S(t)$ is not a Gibbs measure for all $t>t_2$.
\item For $d\geq 3$, if $0<h<h(\beta)$ and $0<\delta<\delta (\beta,h)$,
then there exists $t_3(\beta,h,\delta),t_4(\beta,h,\delta)$ such that
$\nu_{\beta,h} S(t)$ is not a Gibbs measure for all $t\in [t_3,t_4]$.
\end{enumerate}
\et

\bpr
\begin{enumerate}
\item This a consequence of Theorem 4.1. 
\item  This is proved in exactly the
same way as the corresponding point in Theorem \ref{Isingthm}.
\item 
Here we cannot rely on monotonicity as was the case in Theorem \ref{Isingthm}. It is therefore not sufficient to show
that for the fully alternating configuration $\eta^a$, the Hamiltonian
$H(\cdot,\eta^a)$ exhibits a phase transition. We have to show the
following slightly stronger fact: if $m^+_\la (d\sigma)$ is any
Gibbs measure corresponding to the interaction $H(\cdot,\eta^a_\la+_{\la^c})$, then
\be{magn}
\int m^+_\la (d\sigma ) \si (0) >\gamma >0.
\ee
This proof of this fact relies on Pirogov-Sinai theory for the Hamiltonian
$H_t (\cdot,\eta^a_\la +_{\la^c})$.
The first step is to prove that the all-plus-configuration is the unique ground
state of this Hamiltonian. Since the Ising Hamiltonian satisfies the Peierls
condition, we conclude from \cite{Enter} Proposition B.24 that the set of ground states
of $H_t (\cdot,\eta^a_\la +_\la^c)$ is a subset of $\{ +,- \}$. If we drop
the term $H^\delta_t (\cdot,\eta^a_\la+_{\la^c} )$ (i.e., if $\delta=0$), then the
remaining Hamiltonian has as the unique ground state the all-plus-configuration and
satisfies the Peierls condition. Therefore, for $\delta$ small enough, we conclude
from \cite{Enter} Proposition B.24 that $H_t (\cdot,\eta^a_\la +_{\la^c})$ has the
all-plus-configuration as the only possible ground state. From (\ref{pirham}) it is easy
to verify that the all-plus-configuration is actually a ground state for $\delta$
small enough. In order to conclude that for $\beta$ large enough, the unique phase
of $H_t (\cdot,\eta^a_\la +_{\la^c})$ is a weak perturbation of the all
plus configuration (uniformly in $\Lambda$), we can rely on the theory developed in \cite{BKU}, or \cite{DFF} which allows exponentially decaying
perturbations of a finite range interaction satisfying the Peierls condition
(see e.g.\ equations (1.3),(2.2) of \cite{BKU}). Similarly, $H_t (\cdot,\eta^a_\la -_{\la^c})$ has a unique phase which is a weak perturbation of the all minus
configuration. This is sufficient to conclude that no version
of the conditional probabilities is continuous at $\eta^a$, see the discussion \cite{Enter} p.\ 980-981.

\item  We can use the same argument as developed in \cite{Enter},
section 4.3.6, introducing a random perturbation of
the alternating configuration to ``compensate the
uniform magnetic field" (since this requires analysis
of the random field Ising model, we have the restriction $d\geq 3$). The only complication is the extra term in the Hamiltonian
arising from $\delta\not= 0$. This requires Pirogov Sinai theory for
the interaction $H_t (\sigma,\eta)$, where $\eta=\eta^{\epsilon}$ is a random modification
of the fully alternating configuration obtained by flipping the spins
in the alternating configuration with probability $\epsilon/2\beta$
for a flip from $+$ to $-$. Since the couplings between $\eta$ and
$\sigma$ are not finite range, we cannot apply directly
Theorem B31 of \cite{Enter} for the random Hamiltonian $H_t (\sigma, \eta^\epsilon)$.
However, as the interaction decays exponentially fast and Pirogov-Sinai analyses do not distinguish between finite range and exponentially decaying interactions, similar arguments as those developed in
\cite{Zahradnik} still work in our case and yield the analogue
of Theorem B31 of \cite{Enter}. However we have not written out the
details.
\end{enumerate}
\epr
\medskip
\begin{flushleft}
{\bf Remark:}
\end{flushleft}
A result related to Theorem \ref{Dobr*} was obtained in \cite {MV93}.
Although the abstract of that paper is formulated in a somewhat ambiguous
manner, its results apply only to initial measures which are product measures (in particular Dirac measures) .
In particular this includes the case $T_\nu = 0$
and $1<<T_\mu <\infty$.
The results of \cite{MV93} (or \cite{MN}) then imply that
the measure
is Gibbs for all $t>0$. This seems surprising, because $t_2(\beta,\delta)\downarrow 0$ as $\beta\uparrow\infty$. It is therefore better for the intuition to
imagine a Dirac-measure as a product measure than to view it as a limit of
low-temperature measures.


\section{Discussion}
\label{S7}

\subsection{Dynamical interpretation}
\label{S7.1}

In the case of renormalization-group pathologies, the interpretation of
non-Gibbsianness is usually the presence of a {\it hidden} phase transition in
the original system conditioned on the image spins (the constrained system).
In the context of the present paper, we would like to view the phenomenon
of transition from Gibbs to non-Gibbs as a
{\it change in the choice of
most probable
history of an improbable configuration at time $t>0$}.

To that end, let us consider the case of the low-temperature plus-phase of the Ising
model in zero magnetic field ($\beta >>\beta_c$, $h=0$) with an unbiased ($\delta=1$)
infinite-temperature dynamics. Consider the spin at the origin at time $t$
conditioned on a neutral (say alternating) configuration in a {\it sufficiently
large annulus} $\Lambda$ around it. For small times the occurrence of such an
improbable configuration indicates that with overwhelming probability a configuration
very similar was present already at time 0. As the initial measure is an Ising
Gibbs measure, the distribution at time 0 of the spin at the origin is determined
by its local environment only and does not depend on what happens outside the
annulus $\Lambda$. As all spins flip independently, no such dependence can appear
within small times.

However, after a sufficient amount of time (larger than the transition
time $t_2$), if the same improbable configuration is observed, then
it has much more chance of being recently created (due to atypical
fluctuations in the spin-flip processes) than of being the survivor
of an initial state of affairs.  Indeed, to have been there at time
0 is improbable, but to have survived for a large time is even more
improbable. Suppose now that outside the annulus $\Lambda$ we observe
an {\it enormous annulus} $\Gamma$ in which the magnetization is more
negative than $-m^{*} (t)/2$, where $m^{*}(t)$ is the value of the
evolved magnetization (which starts from $m^{*}(0)$ and decays
exponentially fast to zero). Because a large droplet of the minus-phase
shrinks only at finite speed and typically carries a
magnetization characteristic of the evolved minus-phase, with large
probability there was an {\it enormous droplet} of the minus phase (even a bit
larger than $\Gamma$) at time 0, which the spin at the origin
remembers.  Indeed, the probability of this happening is governed by
the size of the {\it surface} of $\Gamma$.  In contrast, the
probability of a large negatively magnetized droplet, arising through a
large fluctuation in the spin-flip process starting from a typical
plus-phase configuration, is governed by the {\it volume} of
$\Gamma$.  Therefore, this second scenario can safely be forgotten.
Although for any size of the initial droplet of the minus-phase there is a
time after which it has shrunk away, for each fixed time $t$ we can choose
an initial droplet size such that at time $t$ it has shrunk no more than
to size $\Gamma$. Since we want the shrinkage until time $t$ to be negligible
with respect to the linear size of $\Gamma$, we need to choose $\Gamma$
larger when $t$ is larger.

Thus, the transition reflects a changeover between two improbable histories
for seeing an improbable (alternating) annulus configuration. It can be viewed as
a kind of large deviation phenomenon for a time-inhomogeneous system.  One could
alternatively describe it by saying that for small times a large alternating
droplet must have occurred at time 0, while after the transition time $t_2$ a
large alternating droplet must have been created by the random spin-flips:
a {\it ``nature to nurture'' transition} \cite{ NS3}. The mathematical analysis
of this interpretation would rely on finding the (constrained) minimum of an
entropy function on the space of trajectories. Alternatively, one could try to
study the large deviation rate function for the magnetization of the measure
at time 0 conditioned on an alternating configuration at time $t$. This rate
function should exhibit a unique minimum for $0\leq t<t_2$ and two minima for
$t>t_2$.

\subsection{Large deviations}
\label{S7.2}

A measure can be non-Gibbsian for different reasons (see \cite{Enter}, section 4.5.5)
One of the possibilities is having ``wrong large deviations", i.e., the
probability
\be{Stsumx}
\nu S(t) \left( \sum_{x\in \la} \tau_x f (\sigma ) \simeq \alpha\right)
\ee
for fixed $t$ and $\alpha\not=\int S(t) f d\nu$ does not decay exponentially
in $|\la|$, i.e., not as $\exp[-|\la| I_f (\alpha )+o(|\la|)]$, or equivalently,
there exists a function $f\in \loc$, $f\geq 0,\ f\not= 0$ such that
\be{Zdfrac}
\lim_{\la\uparrow\Z^d}\frac{1}{|\la|}\log\int\nu S(t)
(d\si)\exp\left[\sum_{x\in\la}\tau_x f(\si)\right] = 0.
\ee
An example where this phenomenon of ``wrong large deviations" occurs
is the stationary measure of the voter model (see e.g.\ \cite{lebscho}).
However, it does not occur in our setting. Namely, if the scale of the
large deviations of the random measure $L_\la = \sum_{x\in \la}\delta_{\tau_x\si}$
under $\nu$ is the volume $|\la|$, then the same holds under $\nu S(t)$ for any $t>0$.
Indeed, by Jensen's inequality and by the translation invariance of the
dynamics we have, for $f\in\loc$, $f\geq 0$, $f\not= 0$,
\beq{Zdfraclog}
&&\limsup_{\la\uparrow\Z^d}\frac{1}{|\la|}\log\int\nu S(t)
(d\si)\exp \left[\sum_{x\in\la}\tau_x f(\si)\right]\nonumber\\
&&\qquad \geq
\limsup_{\la\uparrow\Z^d}\frac{1}{|\la|}\log\int\nu
(d\si)\exp \left[\sum_{x\in\la}\tau_x S(t)f(\si)\right]\nonumber\\
&&\qquad =\sup_\mu \left[ \int S(t)f d\mu - h(\mu|\nu) \right]
\nonumber\\
&&\qquad > 0
\eeq
with $h(\cdot|\cdot)$ denoting relative entropy density. The equality follows from
the volume-scale large deviations of $\nu$, and the last inequality follows
from the fact that $S(t) f\in \ce (\Omega)$, $S(t)f\geq 0$, $S(t)f \not= 0$
imply $\int S(t) f d\nu >0$.
\subsection{Reversibility}
Throughout the whole paper, we have assumed the stationary measure $\mu$
to be reversible. However, this is a condition that only serves to make
formulas nicer. It is not at all a necessary condition: if we consider any high-temperature spin-flip dynamics, then we know that the stationary measure
$\mu$ is a high-temperature Gibbs-measure. Equation (\ref{revRN}) can be
rewritten in the general situation: we have to replace $S_\Lambda (t)$ in the right-hand side by $S^*_\Lambda (t)$, where $S^* (t)$ is the semigroup corresponding to the rates of the reversed process, i.e., the rates
\begin{equation}\label{revrate}
c^* (x,\si) = c(x,\si^x) \frac{d\mu^x}{d\mu}.
\end{equation}
In all the formulas of Section 2, we then have to replace $\E_\si$ by
$\E^*_\si$, referring to expectation in the process with semigroup
$S^* (t)$.
\subsection{Open problems}
\label{S7.3}

\begin{enumerate}
\item
{\bf Infinite-range interactions.}
How much can we save when relaxing the condition that the interactions
be finite-range?
\item
{\bf Trajectory of the interaction.}
In the regime $1<<T_\nu\leq\infty$, $1<<T_\mu\leq\infty$, what can we say about
the trajectory $t\mapsto U_t$? It is not hard to prove that it is analytic in
$\bee_{ti}$ and converges to $U_\mu$. But can we say something about the rate of
convergence? Note that we can view the {\it curve} $\{ U_{\nu_t}: t\geq 0 \}$
as a continuous trajectory in the space $\bee$, interpolating between $U_\nu$
and $U_\mu$, which implies that $\gee$ contains an arc-connected subset. Other
topological characteristics of $\gee$ are discussed in \cite{Enter}, section 4.5.6.
\item
{\bf
Uniqueness
of the transitions.}
Even in the case $T_\mu=\infty$ we have not proved that the transition from Gibbs
to non-Gibbs is
unique
e.g.\ that $t_0 (\beta,0) = t_2 (\beta)$ in Theorem
\ref{Isingthm}. However, we expect that when $h=0$ the alternating configuration
is ``the worst configuration", i.e., the transition is sharp and occurs at the
first time at which the alternating configuration is bad.
\item
{\bf Estimates for the transition times.}
Can we find good estimates for the $t_i$'s as a function of e.g.\ the temperatures,
the magnetic fields and the ranges of the interaction in $\nu$ and $\mu$.
\item
{\bf Weak Gibbsianness.}
In the regimes where $\nu S(t)$ is not a Gibbs measure we expect that we can still
define a $\nu S(t)$-a.s.\ converging interaction $U_t$ for which $\nu S(t)$ is a
``weakly Gibbsian measure" (see \cite{DS2}, \cite{MRV}). This interaction $U_t$ can e.g.\
be constructed
along similar lines as are followed in the proof of
Kozlov's theorem (see \cite{Koz},\cite{MRSV}) and its convergence is
to be controlled by the decay of ``quenched correlations", i.e., the decay of
correlations in the measure at time $0$ conditioned on having a fixed configuration
$\eta$ at time $t$. These correlations are expected to decay exponentially for $\nu S(t)$-a.e.\ $\eta$, which would lead to $\nu S(t)$-a.s.\ convergence of
the Kozlov-potential.
\item
{\bf Low-temperature dynamics.}
The main problem of analyzing the regime $0<T_\mu<<1$ for large $t$ is the impossibility
of a perturbative representation of $-\log p_t (\si,\eta)$.  If we still continue to
work with the picture of the joint Hamiltonian in (\ref{jointham}), then the term
$-\log p_t(\si,\eta)$ will not converge to a $\si$-independent Hamiltonian as
$t\uparrow\infty$. Therefore we cannot argue that for large $t$ the Gibbsianness
of the measure $\nu S(t)$ depends only on the presence or absence of a phase transition
in the Hamiltonian $H_\nu$ of the initial measure $\nu$. The dynamical part of the
joint Hamiltonian can induce a phase transition. The regime $0<T_\mu<<1$ is very
delicate and there is no reason to expect a robust result for general models.
Metastability
will enter. 
\item
{\bf Zero-temperature dynamics.}
What happens when $T_\mu=0$? In this case there is only nature, no nurture. We
therefore expect the behavior to be different from $0<T_\mu<<1$. Trapping
phenomena will enter.
\item
{\bf Other dynamics.}
Do similar phenomena occur under spin-exchange dynamics, like Kawasaki dynamics ?
In particular, how do conservation laws influence the picture (see \cite{dHOO1},
\cite{dHOO2},
\cite{asp})?
\end{enumerate}

{\bf Acknowledgments:} We thank C. Maes and K. Netocny for fruitful
discussions.
A.C.D.v.E. thanks H. van Beijeren for pointing out reference \cite{asp} to him.
Part of this collaboration was made possible by
the ``Samenwerkingsverband Mathematische Fysica''. 
R.\ F.\ thanks the Department
of Theoretical Physics at Groningen for kind hospitality.


\end{document}